\documentclass{aa}
\usepackage{graphicx}
\usepackage{txfonts}
\usepackage{hyperref,textcomp}
\usepackage{ulem}
\hypersetup{
    colorlinks=true,
    linkcolor=blue,
    urlcolor=magenta,
}

\usepackage[dvipsnames]{xcolor}

\begin{document}

      \title{{\it Gaia} GraL: {\it Gaia} gravitational lens systems. IX. Using XGBoost to explore the  \textit{Gaia} Focused Product Release GravLens catalogue}
      
\author{
Quentin Petit\inst{1},
C. Ducourant\inst{1},
E. Slezak\inst{2},
A. Krone-Martins\inst{3,4},\\
C. B\oe hm\inst{5},
T. Connor\inst{6,7},
L. Delchambre\inst{8},
S. G. Djorgovski\inst{9},
L. Galluccio\inst{2},
M. J. Graham\inst{9},
P. Jalan\inst{11},
S. A. Klioner\inst{12},
J. Kl\"uter\inst{13},
F. Mignard\inst{2},
Vibhore Negi\inst{14} ,
S. Scarano Jr\inst{15},
J. Sebastian den Brok\inst{6},
D. Sluse \inst{8},
D. Stern\inst{7},
J. Surdej\inst{8},
R. Teixeira\inst{10},
P.H. Vale-Cunha\inst{10},
D. J. Walton\inst{16},
J. Wambsganss\inst{17}
}

\institute{
\inst{1} Laboratoire d'Astrophysique de Bordeaux, Univ. Bordeaux, CNRS, B18N, all\'ee Geoffroy Saint-Hilaire, F-33615 Pessac, France\\
\inst{2} Universit\'e C\^ote d'Azur, Observatoire de la C\^ote d'Azur, CNRS, Laboratoire Lagrange, Bd de l'Observatoire, CS 34229, F-06304 Nice cedex 4, France\\
\inst{3} Donald Bren School of Information and Computer Sciences, University of California, Irvine, CA 92697, USA\\
\inst{4} CENTRA/SIM, Faculdade de Ci\'encias, Universidade de Lisboa, Ed. C8, Campo Grande, 1749-016, Lisboa, Portugal\\
\inst{5} Sydney Institute for Astronomy, School of Physics, The University of Sydney, NSW 2006, Australia\\
\inst{6} Center for Astrophysics Harvard \& Smithsonian, 60 Garden St., 02138 Cambridge, MA, USA\\
\inst{7} Jet Propulsion Laboratory, California Institute of Technology, 4800
Oak Grove Drive, Pasadena, CA 91109, USA\\
\inst{8} Space sciences, Technologies and Astrophysics Research (STAR) Institute, University of Li\`ege, Belgium\\
\inst{9} Division of Physics, Mathematics, and Astronomy, Caltech, Pasadena, CA 91125, USA\\
\inst{10} Instituto de Astronomia, Geof\'isica e Ci\^encias Atmosf\'ericas, Universidade de S\~{a}o Paulo, Rua do Mat\~{a}o, 1226, Cidade Universit\'aria, 05508-900 S\~{a}o Paulo, SP, Brazil\\
\inst{11} Center for Theoretical Physics, Polish Academy of Sciences, Warsaw, Poland\\
\inst{12} Lohrmann-Observatorium, Technische Universitaet Dresden, D-01062 Dresden, Germany\\
\inst{13} Department of Physics and Astronomy, Louisiana State University, Baton Rouge, LA 70803, USA\\
\inst{14} Inter University Centre for Astronomy and Astrophysics, Post Bag 04, Ganeshkhind, Pune 411007, India\\
\inst{15} Departamento de F\'isica CCET, Universidade Federal de Sergipe,
 Rod. Marechal Rondon s/n, 49.100-000, Jardim Rosa Elze, S\~{a}o Crist\'ov\~{a}o, SE, Brazil\\
\inst{16} Centre for Astrophysics Research, University of Hertfordshire, College Lane, Hatfield, AL10 9AB, UK\\
\inst{17} Astronomisches Rechen-Institut (ARI), Zentrum fur Astronomie der Universitaet Heidelberg (ZAH), Manchhofstr. 12-14, 69120 Heidelberg, Germany\\
\email{quentin.petit.1@u-bordeaux.fr} \\
}
   \date{Accepted February 10, 2025}

  \abstract
   {}
   {Quasar strong gravitational lenses are important tools for putting constraints on the dark matter distribution, dark energy contribution, and the Hubble-Lemaître parameter. We aim to present a new supervised machine learning-based method to identify these lenses in large astrometric surveys. The \textit{Gaia} Focused Product Release (FPR) GravLens catalogue is designed for the identification of multiply imaged quasars, as it provides astrometry and photometry of all sources in the field of 4.7 million quasars.}
   {Our new approach for automatically identifying four-image lens configurations in large catalogues is based on the  eXtreme Gradient Boosting classification algorithm. To train this supervised algorithm, we performed realistic simulations of lenses with four images that account for the statistical distribution of the morphology of the deflecting halos as measured in the EAGLE simulation. We identified the parameters discriminant for the classification and performed two different trainings, namely, with and without distance information.}
   {The performances of this method on the simulated data are quite good, with a true positive rate and a true negative rate of about 99.99\% and 99.84\%, respectively. Our validation of the method on a small set of known quasar lenses demonstrates its efficiency, with 75\% of known lenses being correctly identified. We applied our algorithm (both trainings) to more than 0.9 million quadruplets selected from the \textit{Gaia} FPR GravLens catalogue. We derived a list of 1,127 candidates with at least one score larger than 0.75, where each candidate has two scores—one from the model trained with distance information and one from the model trained without distance information—and including 201 very good candidates with both high scores.}
   {}
   

\titlerunning{XGBoost to search for quads in \textit{Gaia}}
\authorrunning{Q. Petit et al.}
\maketitle

\section{Introduction}\label{intro}

An accurate and unbiased value of the Hubble-Lemaître constant ($H_0$) is key in observational cosmology for characterising the Universe's present-day rate of expansion. Several methods can be used to determine it, and there is currently a tension at a 5$\sigma$ level (e.g. \citealt{2024Wang}) between local measurements involving, for instance, the distances of Cepheids and high redshift ones obtained by fitting the cosmological model to observations of the cosmological microwave background. Unaccounted for biases in the data sets and/or possible inadequacies in the standard $\Lambda$CDM model may explain this tension. Within this context, succeeding in getting more $H_0$ estimates from quasar strong gravitational lenses is of great interest. This approach, first discussed in \cite{1964Refsdal}, relies on the observed time delay for propagating changes in the source brightness between lensed images. It is indeed independent from both cosmic distance ladder determinations and type Ia supernovae, gravitational source detections, and cosmological microwave background analyses, with a final accuracy depending mainly on the ability to model the projected mass distribution of the lenses and the number statistics of the sample. Today, the Hubble-Lemaître constant can be determined this way with a precision of up to 2.4\%, assuming a spatially flat cosmology and accounting for systematic errors \citep{2020Wong}.

The main limiting factor to reach the desirable 1\% level is the small number of quasar gravitational lenses suitable for such studies (only six gravitationally lensed quasars are involved in the above H0LICOW paper). Even before being able to monitor the confirmed lenses on a decade-long term and obtaining the required richly sampled light curves (e.g. the COSMOGRAIL program \citep{2005Courbin,2020Millon}), it is first mandatory to identify systems with two or more lensed images among millions of sources, with the even rarer quadruply imaged quasars (quads) benefiting from finer modelling of the deflector.

The landscape has evolved in recent years with the discovery of dozens of new lensed quasars in large-scale optical surveys such as the Sloan Digital Sky Survey \citep[SDSS;][]{2009SDSS} and the Dark Energy Survey \citep{ 2016DES} thanks to the development and automation of lens identification algorithms.
In that respect, the ESA \textit{Gaia} mission currently plays a considerable role by accelerating the discovery of quads \citep {2018Ducourant,2021Stern}.

The common factor in all of these blind searches in large data sets is the use of powerful methods to sift through the images and automatically select lens candidates. This research has especially motivated the use of artificial intelligence-based strategies, such artificial neural networks (ANNs; \citealt{1957Rosenblatt}) and convolutional neural networks (CNNs; \citealt{1989LeCun}), to analyse first more or less complex simulations of strongly lensed systems for various surveys (e.g. \citealt{2017Hezaveh, 2018Schaefer, 2018Lanusse, 2019Pearson, 2024Leuzzi}) and to then look in parallel for such events in wide-field imaging surveys such as the Canada-France-Hawaii Telescope Legacy Survey \citep[CFHTLS;][]{2017jacobs}, the COSMOS field \citep{2018Pourrahmani},
the Kilo Degree Survey \citep[KiDS;][]{2017Petrillo,2019Petrillo1,2019Petrillo2,2020He,2021Li}, the Dark Energy Survey \citep{2019Jacobs1,2019Jacobs2,2022Rojas, 2023Zaborowski}, the Dark Energy Spectroscopic Instrument (DESI) Legacy Imaging Surveys \citep{2020Huang,2021Huang}, the Panoramic Survey Telescope and Rapid Response System (Pan-
STARRS) survey \citep{2020Canameras}, the VST Optical Imaging of the CDFS and ES1 fields \citep[VOICE survey;][]{2022Gentile}, and the Hyper-Suprime Cam Subaru Strategic Program \citep[HSC-SSP;][]{2024Moskowitz}.

It is important to note that the aforementioned large-scale surveys and studies are predominantly ground-based. Consequently, only gravitational lenses with angular separations between lensed images larger than about 1.5 arcseconds have been detected in practice.
Overcoming this limitation to also detect compact gravitational lenses with angular separations smaller than 1 arcsecond requires a high angular resolution that is easier for space observations to reach, which also have the benefit of a very stable instrumental response.  The ESA \textit{Gaia} space observatory with its all-sky data releases is without equivalent for such a purpose, with an unparalleled theoretical angular resolution of 0.18".

This paper presents a new machine learning-based approach, namely, the use of the eXtreme Gradient Boosting (XGBoost) algorithm, to search for gravitational lenses of quasars in the \textit{Gaia} data releases, especially quads. Since \textit{Gaia} provides catalogues of positions rather than images, it is essential to work at the catalogue level, making supervised machine learning algorithms particularly well suited for this task. Our method is rooted in improved simulations of gravitational lenses and a careful selection of the relevant information for this goal to make significant progress in the blind identification of such systems in very large catalogues.

This paper is structured as follows. Section \ref{sec:FPR} presents the \textit{Gaia} Focused Product Release (FPR) GravLens catalogue, highlighting its relevance and potential for our study. Section \ref{sec:XGBoost} introduces the XGBoost algorithm. In Sect. \ref{sec:training}, we outline the construction of our training set, focusing on the creation of a realistic catalogue of simulated lenses. Section \ref{sec:classification} details the crucial discriminant parameters for the classification task. Section \ref{sec:performances} describes the XGBoost training process, offering insights into its performance metrics and efficacy in the given context. Section \ref{sec:applications} presents the application of our trained model to the GravLens dataset. Section \ref{sec:conclusion} summarises our findings and indicates potential improvements for future work.

\section{The \textit{Gaia} Focused Product Release GravLens catalogue}\label{sec:FPR}

Current releases of \textit{Gaia} data remain incomplete for the lenses of quasars with the smallest angular separations \citep{2017Arenou,2021Fabricius,2021Torra}. One or more lensed images of some known systems have indeed no counterpart in \textit{Gaia} DR3 \citep{2018Ducourant}, although they are detected by the satellite. This situation has slowed the identification of new lenses in \textit{Gaia} data because most of the as yet undiscovered gravitational lenses of quasars are characterised by small angular separations. To address this, the \textit{Gaia} Data Processing and Analysis Consortium (\textit{Gaia} DPAC) has developed a dedicated processing chain aimed at analysing the environment around quasar candidates and producing a catalogue of sources near these candidates. This catalogue is more complete at smaller separations compared to \textit{Gaia} DR3.

 This chain uses an unsupervised clustering algorithm widely used in machine learning and data analytics to cluster raw \textit{Gaia} measurements around quasars within 6 arcseconds. This so-called density-based spatial clustering of Applications with noise (DBSCAN) algorithm \citep{1996ester} groups the individual epoch detections in right ascension and declination coordinates with angular separations smaller than a given threshold, allowing new sources not previously published in current \textit{Gaia} catalogues to be identified. The whole set of sources found in the neighbourhood of a quasar is called hereafter a multiplet.
 
 This chain works with raw data so that the astrometry and the photometry that it produces are less accurate than those of the \textit{Gaia} DR3. The related GravLens catalogue with the astrometry and the photometry of all detected sources is presented in the FPR publication of \textit{Gaia} \citep{2023GaiaFPR}. It includes 3\,760\,032 investigated quasars and a total of 4\,760\,920 sources detected in their vicinity (including the quasars themselves). This catalogue is enriched by $\sim$103\,000 new sources not present in \textit{Gaia} DR3.

\begin{figure}
    \centering
    \includegraphics[width=0.36\textwidth]{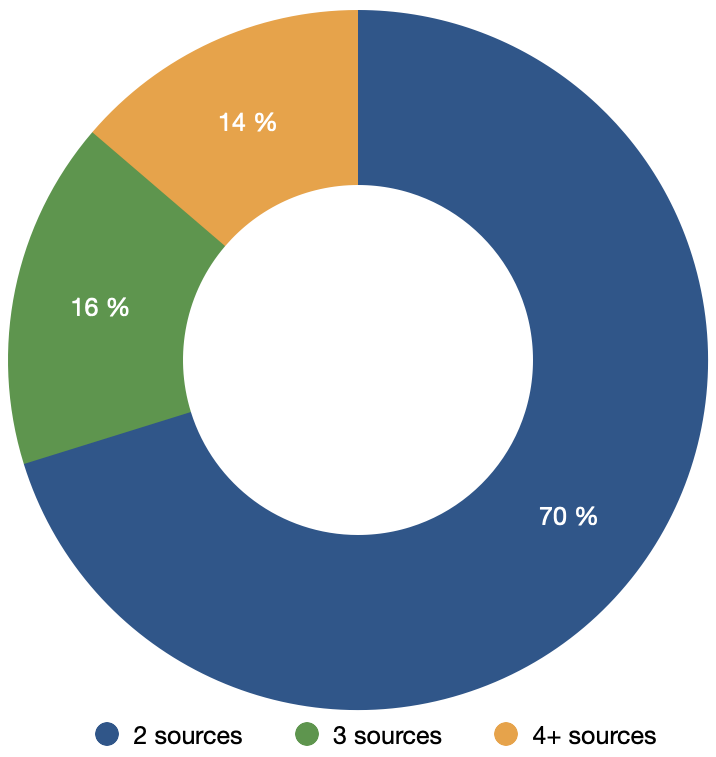}
    \caption{Distribution of the number of sources contained in the 501\,380 GravLens multiplets with more than one component.}
    \label{multiplets}
\end{figure}

 In the GravLens catalogue, 87\% of the quasars are single sources, and neighbouring sources are detected around the quasar in 501\,380 cases. The number of sources found in these multiplets is illustrated in Fig. \ref{multiplets}. Most (70\%) of the multiplets are composed of two sources. The three-source multiplets and the four-or-more source multiplets concern 15\%  and 14\% of the cases, respectively. 

We used this catalogue to search for quads and focused the application of the algorithm we developed on multiplets consisting of four sources or more. Of course, lenses can also be found in multiplets with three sources (one of the images of a quad may not be detected by \textit{Gaia} because it is too faint given \textit{Gaia}'s magnitude limit of approximately G = 21). Analysing the lenses was the next step in our study.

\section{The XGBoost algorithm to search for lenses}\label{sec:XGBoost}

The search for quads in very large data sets such as the GravLens catalogue imposes the use of machine learning techniques. We chose a method based on supervised learning leveraging ensemble machine learning techniques in order to improve prediction accuracy compared to a single model. This type of algorithm is less prone to produce results excessively influenced by specific training data or minor variations in input data, and its predictions are therefore more stable and reliable in different conditions or when encountering variations in these data. This approach also helps reduce overfitting and provides robust results.

To explore the extensive data sets released by \textit{Gaia}, we relied on the machine learning method XGBoost \citep{2016Chen} for the lens recognition process. XGBoost is an algorithm that combines ensemble learning with decision trees to create a robust predictive model. Thanks to its capacity to capture intricate relationships between input variables, XGBoost excels in data classification and is especially well suited for high-dimensional problems. It operates by training a sequence of successive decision trees. Each tree is added to the ensemble iteratively with the aim of enhancing the prediction accuracy of the model under construction. At each iteration, the model predicts the residuals (the disparity between the current predictions and the true values) rather than the raw values themselves. This approach diminishes the residual error at each step, enhancing the model's accuracy over time. Decision trees are constructed to minimise the loss function and integrate regularisation techniques to prevent overfitting.

The XGBoost model performs better than extremely randomised trees (ERT;  \citealt{2006geurts}) when dealing with class imbalance, which is the case in our application since only one over 1000 quasars is expected to be lensed, and one-fifth of them are expected to be a quad. The boosting algorithm learns iteratively from the errors of the previous tree. Therefore, if a tree fails to predict a particular class (often the imbalanced one), the subsequent tree will assign more weight to this sample. Essentially, this process aims to balance the model by prioritizing underrepresented categories. In contrast, the ERT algorithm lacks a mechanism to address data imbalance.

\section{A realistic training set}\label{sec:training}

To construct the training dataset for XGBoost, we set up two classes of objects. The first class contains gravitational lenses, while the second class consists of groups of stars. 
We intended to produce a realistic training set for our algorithm essentially by improving the simulations of lenses representing the first class of sources. For the second class of sources, we used the star clusters derived by \cite{2019Delchambre}, as they are a good representative of \textit{Gaia}'s stellar populations.

\subsection{First class: Simulations of realistic gravitational lenses}
There are less than 90 spectroscopically confirmed quads, and this severely limits the creation of a comprehensive labelled catalogue encompassing all potential configurations \citep{2018Ducourant}. To address the scarcity of known gravitational lenses of quasars, one can instead use simulations to train the classification algorithms, as done by \citep{2019Delchambre} who trained a model based on ERT with simulations. However, these simulations were produced using a uniform distribution of parameters describing the morphology and velocity dispersion of the deflecting galaxies due to the unavailability, at that time, of more precise data. As a result, the produced simulations contained a significant proportion of non-realistic configurations, leading to a classification with an excessively high rate of false positives.
This emphasises the importance of having a highly realistic training set, as it directly impacts the effectiveness and reliability of the model to identify gravitational lenses. To explore the extensive data sets of GravLens with XGBoost, we created numerous gravitational lens simulations using non-uniform distributions of the lens parameters as measured by \cite{2023Petit} on the cosmological EAGLE simulations \citep{2015Schaye} and including a realistic population of quasars. 

\subsubsection{Background sources: Quasars}
\begin{figure}
    \centering
    \includegraphics[width=0.25\textwidth]{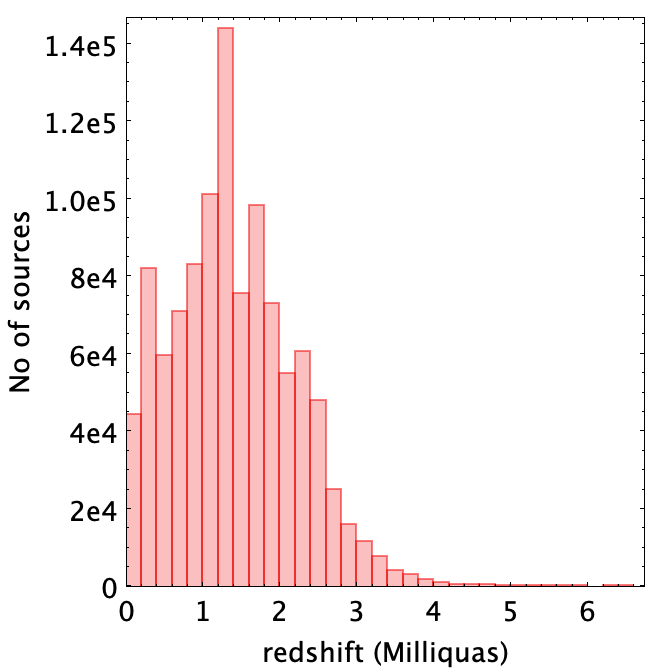}\includegraphics[width=0.25\textwidth]{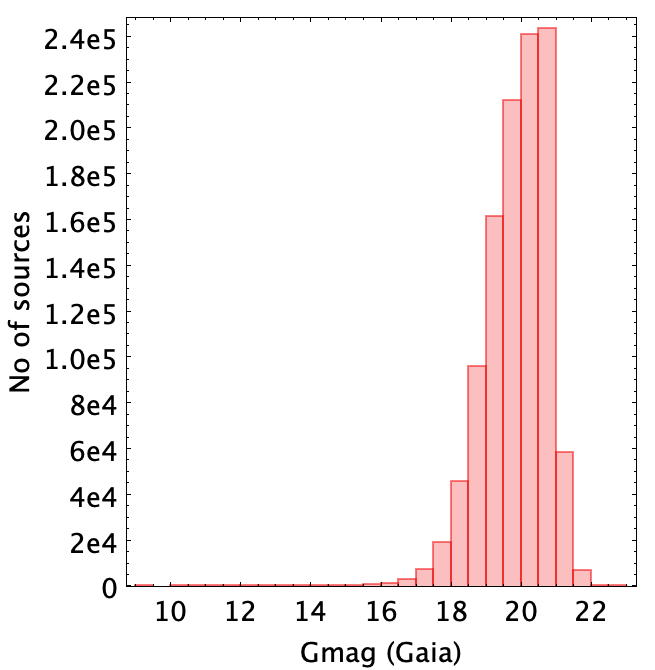} 
    \caption{Distribution of redshifts and \textit{Gaia} G magnitudes for the quasars in common between Milliquas and GravLens.}
    \label{miliquas_distribution}
\end{figure}

We used the Million Quasars Catalog \citep[Milliquas;][]{2021Milliquas} for the simulation of a realistic population of quasars. In this catalogue, 864\,000 quasars have a redshift measurement and an entry in the GravLens catalogue. Figure \ref{miliquas_distribution} shows the distribution of their redshifts and G magnitude. The distribution of redshifts peaks around $z=1.5$ and extends up to $z=6$. The median G-band magnitude of the sample is 20.

\subsubsection{Lenses: Galaxies from the EAGLE simulation}

In a recent paper \citep{2023Petit}, we analysed the properties of galaxies from the hydrodynamic EAGLE simulations \citep{2015Schaye}. Specifically, we measured the ellipticity of galaxies projected onto the plane of the sky, their half-mass radius, and their velocity dispersion ($\sigma_v$), and we collected redshifts and masses. This provided us with statistical distributions of parameters characterising the deflecting galaxies. Our aim was to generate realistic lens simulations by utilising these statistical distributions as priors.

\subsubsection{The lens model}\label{model}

We simulated gravitational lensing phenomena using a singular isothermal ellipsoid (SIE) model \citep{1994Kormann}. The SIE model expands upon the singular isothermal sphere (SIS) model by incorporating ellipticity, thus providing a more versatile representation of elliptical galaxies as gravitational lenses. The lensing potential of the SIE model is expressed as $\Phi (x, y) = \theta_E \sqrt{q^2 x^2 + x^2 / q}$, where $\theta_E$ denotes the Einstein radius defining the strength of the lensing effect, $x$ and $y$ represent coordinates of the background source in the lens plane, and $q$ is the axis ratio of the deflecting galaxy.

\begin{figure}[!h]
    \centering
    \includegraphics[width=0.49\textwidth]{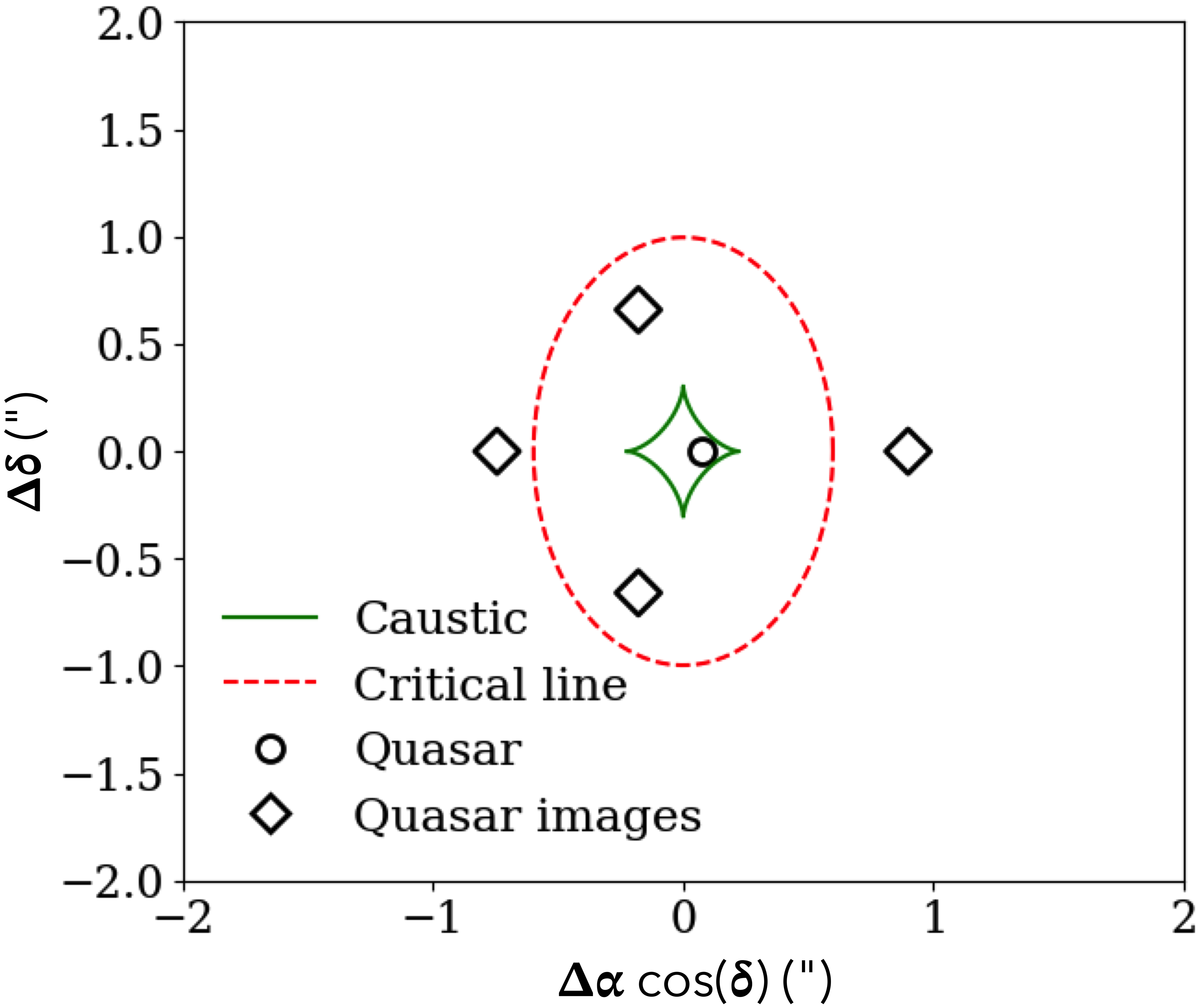}
        \caption{Projected sky coordinates of a typical gravitational lens system obtained with an SIE model with a quasar placed at $z=1.0$ and a lens at $z=0.5$ with $q=0.6$. The background quasar is placed inside the green curve so that the lens produces four distinct images of the quasar. }
    \label{caustic}
\end{figure}

Figure \ref{caustic} illustrates a typical gravitational lens system obtained with an SIE model featuring a quasar at $z=1.0$ and a lens at $z=0.5$ with $q=0.6$. The plot is centreed at the galaxy's centre. The green curve (diamond shape) represents the caustic line in the quasar plane, delineating the boundary between regions where light rays converge to form multiple images and regions where they do not. When the quasar lies on this green line, gravitational lensing magnification formally becomes infinite, resulting in highly distorted and amplified images. The red dotted line represents the critical line in the lens plane, which marks the boundary between areas where light is deflected inward to form multiple images and areas where it is deflected outward without forming multiple images.

\subsubsection{Calculation of the Einstein radius}
One of the quantities that characterises a gravitational lens is its Einstein radius, a physical measure of the angular scale of the phenomenon. The Einstein radius of the quasar plus lens pair is calculated for an SIE model using the relation

\begin{equation}\label{TE}
\theta_E = 4 \pi \left( \frac{\sigma_v}{c} \right)^2 \frac{D_{LS}}{D_S},
\end{equation}

\noindent where $\sigma_v$ is the velocity dispersion of the deflector, $D_{LS}$ is the angular diameter distance between the deflector and the source, and $D_S$ is the angular diameter distance from the observer to the source.

The virial theorem states that the time-averaged kinetic energy of a system is equal to half the time-averaged potential energy. By applying this theorem to a relaxed gravitational system, we can express the velocity dispersion (\(\sigma\)) as a function of the system's mass (\(M\)) and a characteristic radius (\(R\)):

\begin{equation}
\sigma = \sqrt{\frac{3}{5}\frac{G}{R}M}.
\end{equation}

The choice of the characteristic radius is critical and should be representative of the size or extent of the system's projected mass distribution. One common choice is the half-mass radius (\(R_{\text{hm}}\)), which corresponds to the radius within which half of the total mass of the system is included. Utilising the half-mass radius (\(R_{\text{hm}}\)), we can estimate the velocity dispersion (\(\sigma\)) of the halos in the simulation based on the mass (\(M_{\text{hm}}\)) within that radius:

\begin{equation}
\sigma = \sqrt{\frac{3}{5}\frac{G}{R_{\text{hm}}}M_{\text{hm}}}.
\end{equation}

\begin{figure}[!h]
    \centering
    \includegraphics[width=0.49\textwidth]{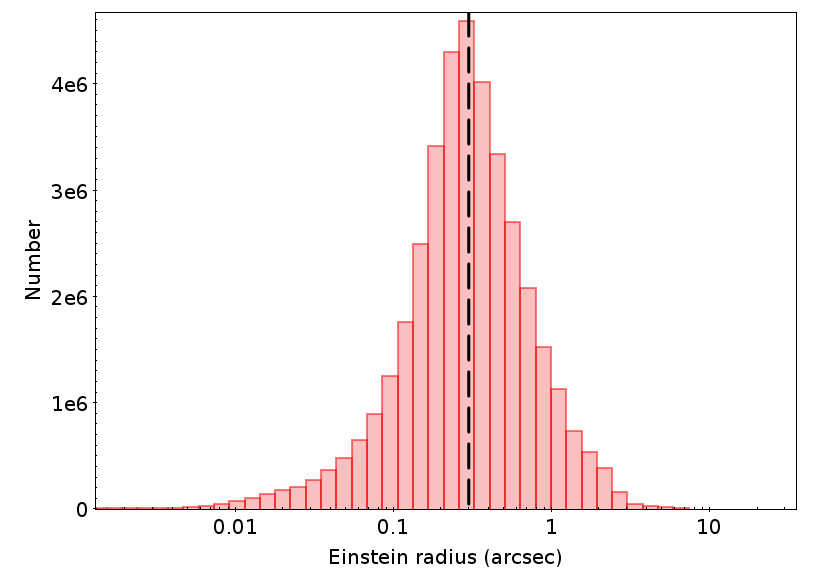}
        \caption{Distribution of Einstein angular radii in logarithmic scale obtained by combining the distribution of quasars from Milliquas with the EAGLE galaxies. Dotted line corresponds to the limit of \textit{Gaia}'s GravLens resolving separation (0.3").}
    \label{plot:histo_einstein_radius}
\end{figure}

Given the vast number of possible combinations between the 340\,719 EAGLE halos analysed and the 864\,000 Milliquas quasars with redshift measurement, it is impractical to calculate all Einstein radii. To obtain a realistic distribution of Einstein radii ($\theta_E$), we adopted an approach in which we randomly selected 500 quasars from our list to be placed behind each EAGLE halo. These 500 quasars were chosen to match the redshift and magnitude distributions of the initial sample from the distribution of redshift in EAGLE simulation snapshots. This method allowed us to calculate 170\,359\,500 $\theta_E$ radii. The distribution of these Einstein radii is presented in Fig. \ref{plot:histo_einstein_radius}. 

We observed that many radii are extremely small and thus correspond to configurations that the \textit{Gaia} satellite will not resolve. For our training set, we selected the simulations with an Einstein angular radius larger than 0.3", corresponding to \textit{Gaia}'s resolving power, since closer sources are merged into single sources in the GravLens catalogue \citep{2023GaiaFPR}.

\subsubsection{Solving the lens equation}
In gravitational lensing, the Einstein radius is typically normalised to unity, which means it is scaled to a standard reference size. This normalisation simplifies the lens equation by reducing the complex geometric relationships to a standardised configuration. However, in the present work, we wanted to keep full track of the astrometry of the quads and did not want to work with these normalised configurations. The Einstein radius allowed us to denormalise the configurations and ascertain the actual sizes of the lenses. To generate a set of simulations, we simulated various lensing configurations by randomly selecting the position of the quasar within the diamond caustic structure that is typical of an elliptical potential (SIE), $\Phi$, as defined in \ref{model}.

\subsubsection{Shear}
\begin{figure}[]
\centering
    \includegraphics[width=0.50\textwidth]{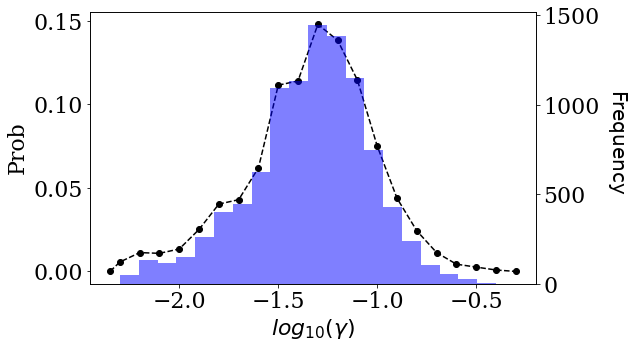}
        \caption{Distribution of shear from the study by \cite[dotted line]{2003Holder} and the random selection of $N = 10,000$ shear values respecting this distribution (blue histogram).}
    \label{shear_distribution}
\end{figure}

An important aspect of our simulations is the inclusion of external shear to account for the superposition of masses along the line of sight as well as for the presence of other masses nearby at the same redshift as the main lens. The shear utilised in this work is based on the distribution estimated by \cite{2003Holder} derived from the public simulations of the Semi-Analytic Galaxy Formation - GIF project \citep{1999Kauffmann}. Figure \ref{shear_distribution} illustrates the shear distribution adopted for our study. Taking a shear into account has a major impact on the astrometry and photometry of the images produced by the lens and is essential for carrying out realistic simulations of gravitational lenses.

\subsubsection{Accounting for astrometric errors}
The astrometric errors in the positions of quasars have a significant impact on the astrometry and photometry of the image configurations generated when solving the lens equation. Therefore, we introduced a Gaussian noise representative of GravLens errors (60 mas on positions and 0.15 mag on magnitudes) on the positions and magnitudes of the quasar images. This step enabled us to produce configurations that better reflect actual observational conditions and capture the fluctuations and inaccuracies inherent in genuine astronomical observations. Adding these uncertainties also allowed the method to take into account part of the effects induced on the luminosity of the images by microlensing.

\subsection{Second class: Stellar multiplets}
\textit{Gaia} primarily observes stars, so most of the multiplets in the GravLens catalogue consist of groups of stars. Therefore, it was essential to include a large number of stellar multiplets in our training set so that XGBoost could learn to distinguish them from images of multiply imaged quasars.

We utilised the stellar multiplets isolated in \textit{Gaia} data by \cite{2019Delchambre}. From these, we extracted 65,693 multiplets, each comprising four stars. Each source is characterised by its equatorial coordinates, \textit{Gaia} $G$ magnitude, and errors. We converted the celestial coordinates (ra, dec) to Cartesian coordinates (x, y) using gnomonic projection.

\subsection{The training catalogue}
To train the XGBoost algorithm for an optimal classification of gravitational lens configurations in various situations, we set up a training catalogue that includes  44\,339 realistic lens simulations with external shear and characteristic images with a separation larger than 0.3" and 65\,332 stellar multiplets. The two classes of objects are balanced in number. 

This training catalogue could be improved in future studies. Our simulations are based on halos from the EAGLE simulations. The analysis was performed on the `small' EAGLE simulation, which contains only a few massive halos with masses greater than $10^{12}$ M$_\odot$, which are the ones likely to produce large configurations of lensed quasars. Consequently, our resulting set of lens simulations contains a small number of large configurations. To overcome this limitation, one could artificially add more massive halos to our list of lensing galaxies or analyse larger EAGLE simulations. 

\section{Discriminant parameters for classification}\label{sec:classification}

\subsection{Basic parameters}
We needed to define the parameters that the algorithm would use for classification. The choice of these parameters is crucial for optimising the performance of the model. Ideally, the parameters should be concentrated in a low-dimensional subspace distinct from the others. 
The GravLens catalogue only provides equatorial coordinates and \textit{Gaia} G band magnitudes for each source of the multiplets. Our objective in this section is to find an efficient parameter space that facilitates the discrimination between quadruplets of lensed quasars and random configurations of four stars.  

First, we computed the luminosity ratio between each pair of sources in the quadruplets and ranked the four sources according to their respective amplification (or relative flux) with labels A to D in descending order. Then, we calculated the Euclidean distances $d_1$ to $d_6$ between the four sources in the projected plane and recorded the minimum (MinDist) and maximum (MaxDist) values. We also calculated all the angles for each trio of images. Thus, we obtained twelve distinct angles: $\widehat{ABC}$, $\widehat{BAD}$, $\widehat{ADC}$, $\widehat{BCD}$, $\widehat{ABD}$, $\widehat{CBD}$, $\widehat{BAC}$, $\widehat{DAC}$, $\widehat{BDA}$, $\widehat{BDC}$, $ \widehat{BCD}$, $\widehat{ACB}$, $\widehat{DCA}$. 

Finally, to achieve uniform scaling across the configurations, we normalised the distances to the maximum distance found. This preserves the relative spatial relationships within each configuration while allowing for comparison between multiplets.

\subsection{A new reference plane}
The multiplets of stars are random configurations, whereas the images of a lens follow a certain order. This is why we are looking at lens simulations and searching for the combination of parameters that will allow us to distinguish them from random configurations. To ensure uniformity and facilitate the comparison of multiplets, we systematically centred each configuration on its brightest source (A component) and rotated it so that the second brightest source (B) is aligned along the vertical axis, thus creating a new reference plane ($X_1,X_2$).

\begin{figure}[!h]
\centering
    \includegraphics[width=0.49\textwidth]{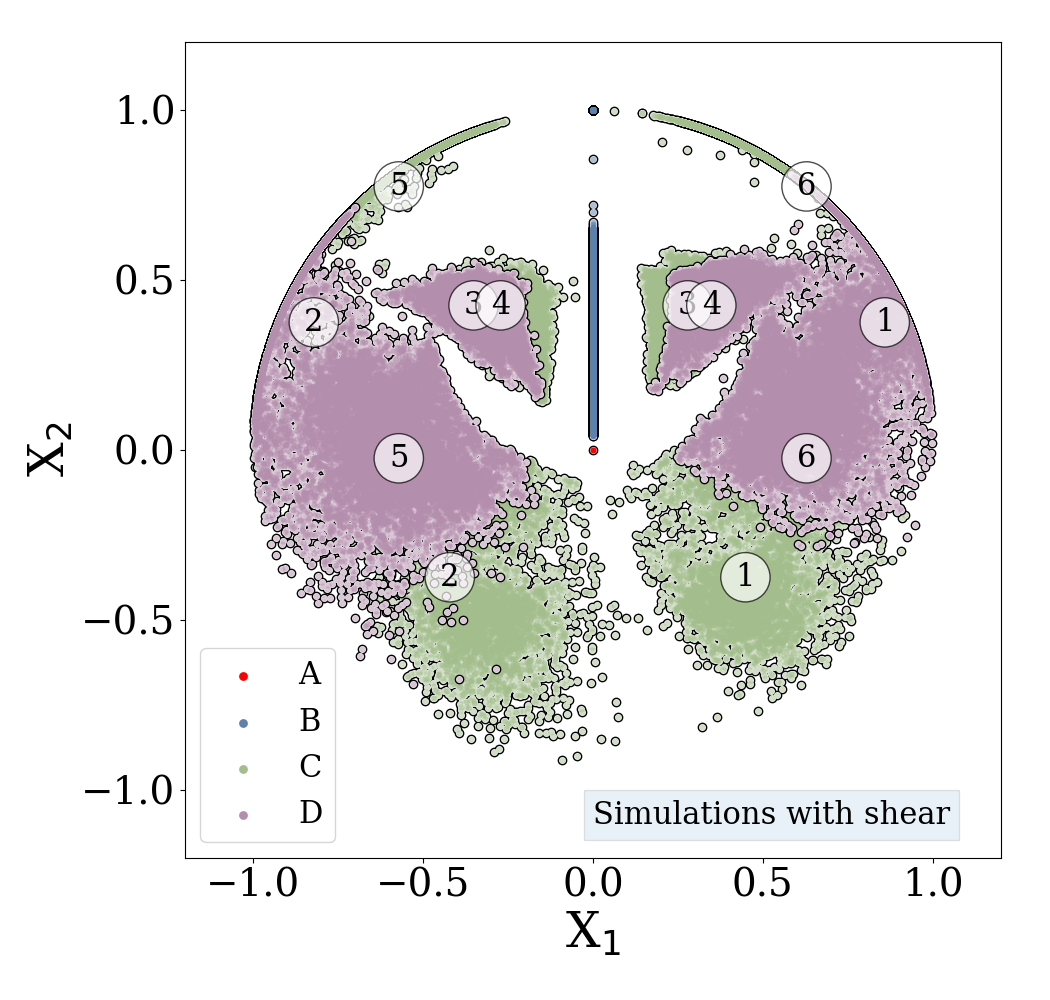}
    \caption{Distribution of images A,B,C and D of the set of simulations of gravitational lenses in the ($X_1,X_2$) plane. Red central dot corresponds to images A, vertical blue line to images B, green zones to images C and pink zones to images D. }
    \label{zones}
\end{figure}

We present in Fig. \ref{zones} the distribution in the ($X_1,X_2$) plane of images A, B, C, and D of the set of simulations of gravitational lenses that we performed after the normalisation and reorientation steps. We observed that the images of gravitational lenses fall into specific and well-separated regions coloured respectively in red (A), blue (B), green (C), and pink (D). In the figure, we label the zones containing C and D from 1 to 6 (for example, when C is in the green zone 2, D is in pink zone 2). The pink and green zones 3 and 4 (and inversely) overlap.

\begin{figure}[!h]
\centering
    \includegraphics[width=0.49\textwidth]{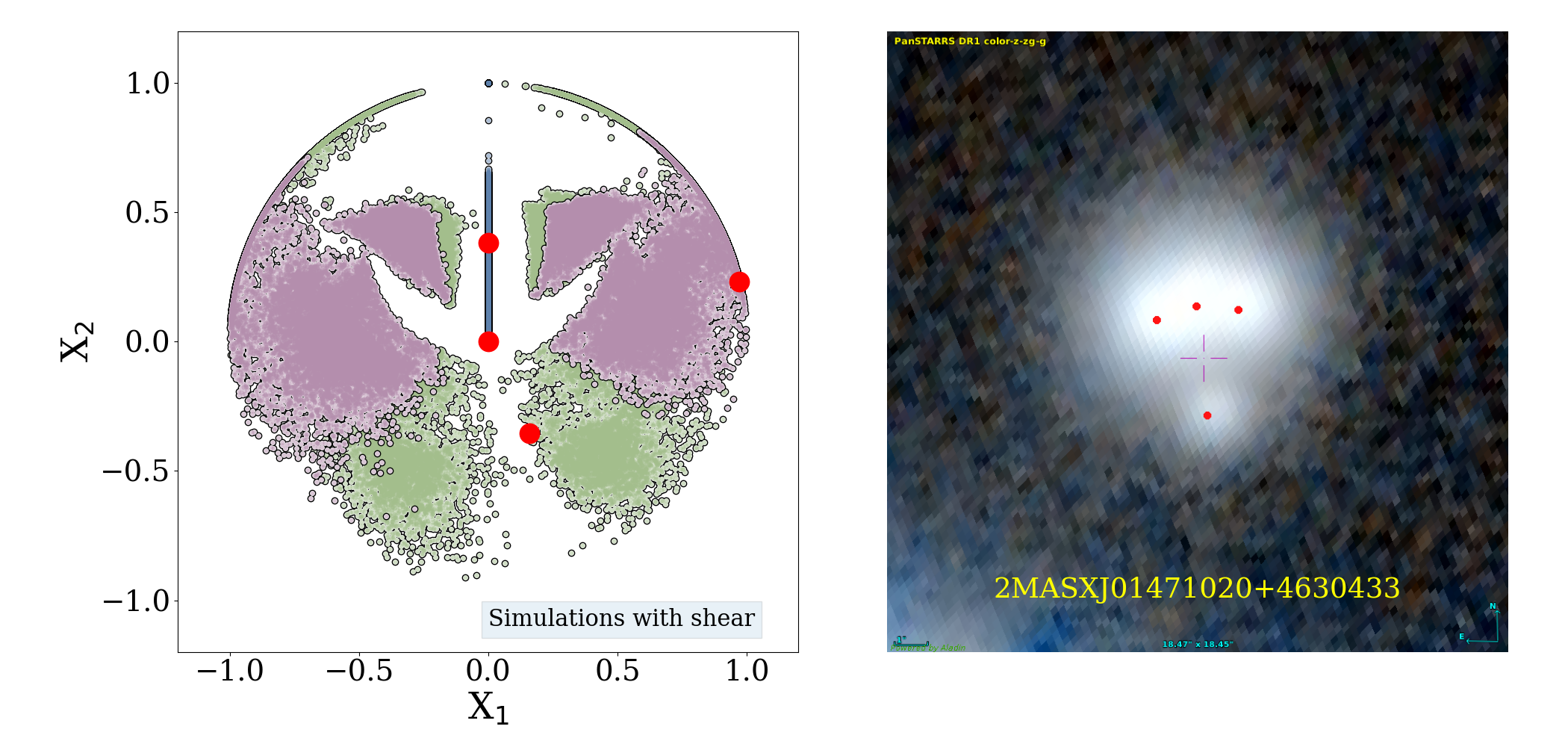}
    \caption{Left panel: Distribution of images of the typical cusp configuration of J014710+463040 in the ($X_1,X_2$) plane. Right panel: Pan-STARRS image of the gravitational lens J014710+463040.}
    \label{cusp_config}
\end{figure}

In this complex figure, we observed three types of configurations. The first one corresponds to the cases where the B component is at coordinates (0, 1), blue point, and the C and D images are both located in zones 3 or 4 and correspond to ‘Einstein cross’ type configurations. In the two other configurations, B is closer to A (lying along the vertical blue line) and C and D are on the same side of the plot concerning the vertical axis, in the external pink and green regions (1, 2, 5, or 6). This type of configuration is illustrated in Fig. \ref{cusp_config}, which presents the J014710+463040 gravitational lens configuration in the ($X_1, X_2$) plane (red dots) over-plotted on top of the different zones identified. In that case, the C and D images both lie on the left part of the plot in zone 2. 

\begin{figure}[!h]
\centering
    \includegraphics[width=0.49\textwidth]{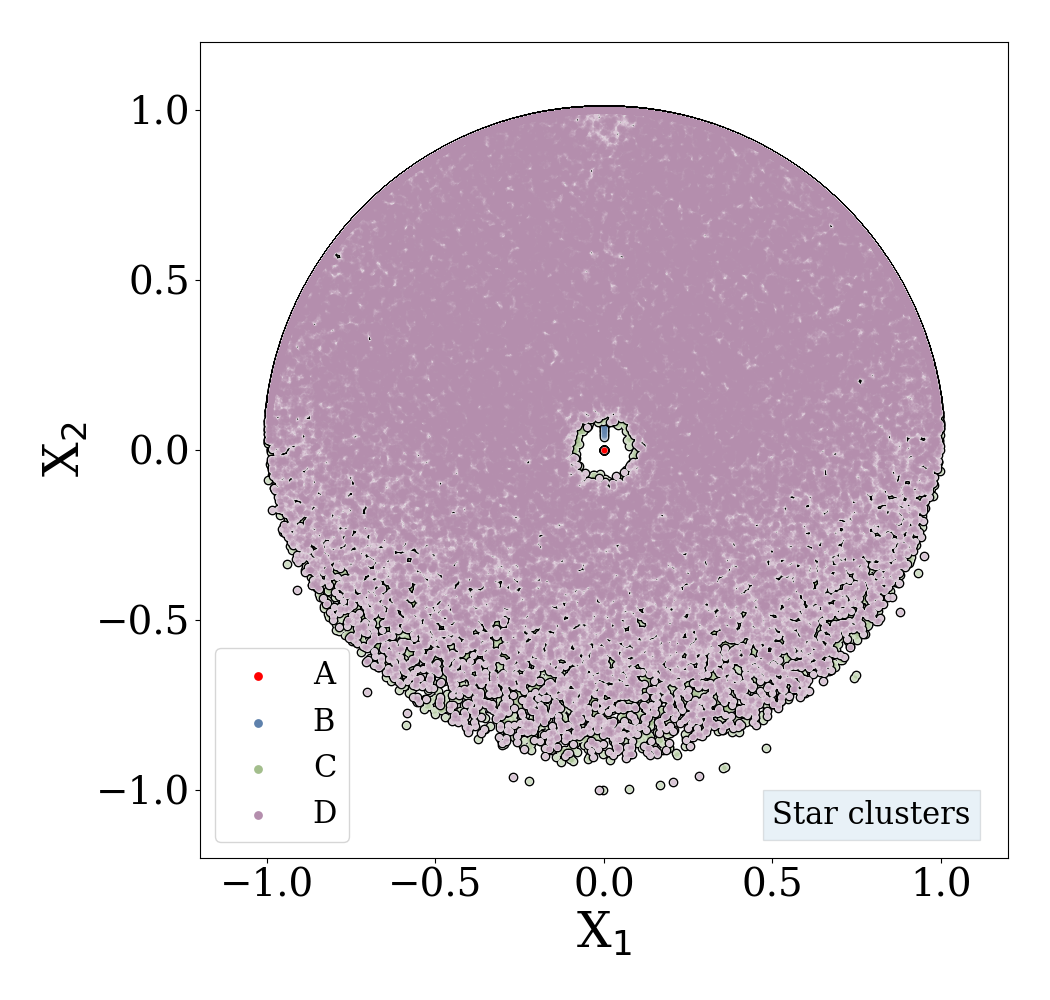}
    \caption{Distribution of images of star clusters from \textit{Gaia} in the ($X_1, X_2$) plane.}
    \label{cerveau_stars}
\end{figure}

As one can see, the organisation of the images into specific zones and at specific angles in the ($X_1,X_2$) plane is crucial information for proper separation between lenses and groups of stars. Indeed quadruplets of stars do not show any specific pattern in the ($X_1,X_2$) plane, as seen in Fig. \ref{cerveau_stars}, which presents the distribution of groups of four stars from \textit{Gaia} in the ($X_1,X_2$) plane. 

\section{XGBoost training}\label{sec:performances}

\subsection{Parameters}\label{params}
Table \ref{table_parametres} presents the list of parameters we collected to train XGBoost. We trained XGBoost with this list of parameters and examined the feature's importance. This measure helped determine which features impact the predictions the most and can therefore be considered the most informative for the model. Feature importance is calculated by XGBoost using different methods, such as how often a feature is used when building decision trees and the average split score improvement achieved from that feature.

\begin{table*}
\centering
\caption{Training parameters selected for the XGBoost classification algorithm.}
\begin{tabular}{ll}
\hline
Parameters & Description \\
\hline
$(X_1, X_2)^i$, (i=A,..D) & Coordinates of images in the ($X_1$, $X_2$) plane \\
$\widehat{ABC}$, ..., $\widehat{DAC}$  &Set of 12 angles between the four images in the $(X_1, X_2)$ plane\\
Nmu1, ..., Nmu4     &Flux ratios to flux(A) \\
\hline
d1, ..., d6         &Distances between the four images \\
Nd1, ..., Nd6       &Normalised distances by MaxDist \\ 
MaxDist             &Maximum distance between images \\
MinDist             &Minimum distance between images \\
\hline
\end{tabular}
\label{table_parametres}
 \tablefoot{
 The first set of parameters was used in training(basic), and the entire set of parameters was used in training(dist).
 }
\end{table*}

We carried out two separate training sessions, one using the distance parameters, training(dist), and the other not using them, training(basic). The first session, training(dist), used all the parameters listed in Table \ref{table_parametres}, while the second session, training(basic), used only the $(X_1, X_2)^i$ (i=1,4) positions of each image, the angles ($\widehat{ABC}$, ..., $\widehat{DAC}$), and the flux ratios relative to image A (Nmu1, ..., Nmu4).

\begin{figure*}[!h]
\centering
    \includegraphics[width=0.42\textwidth]{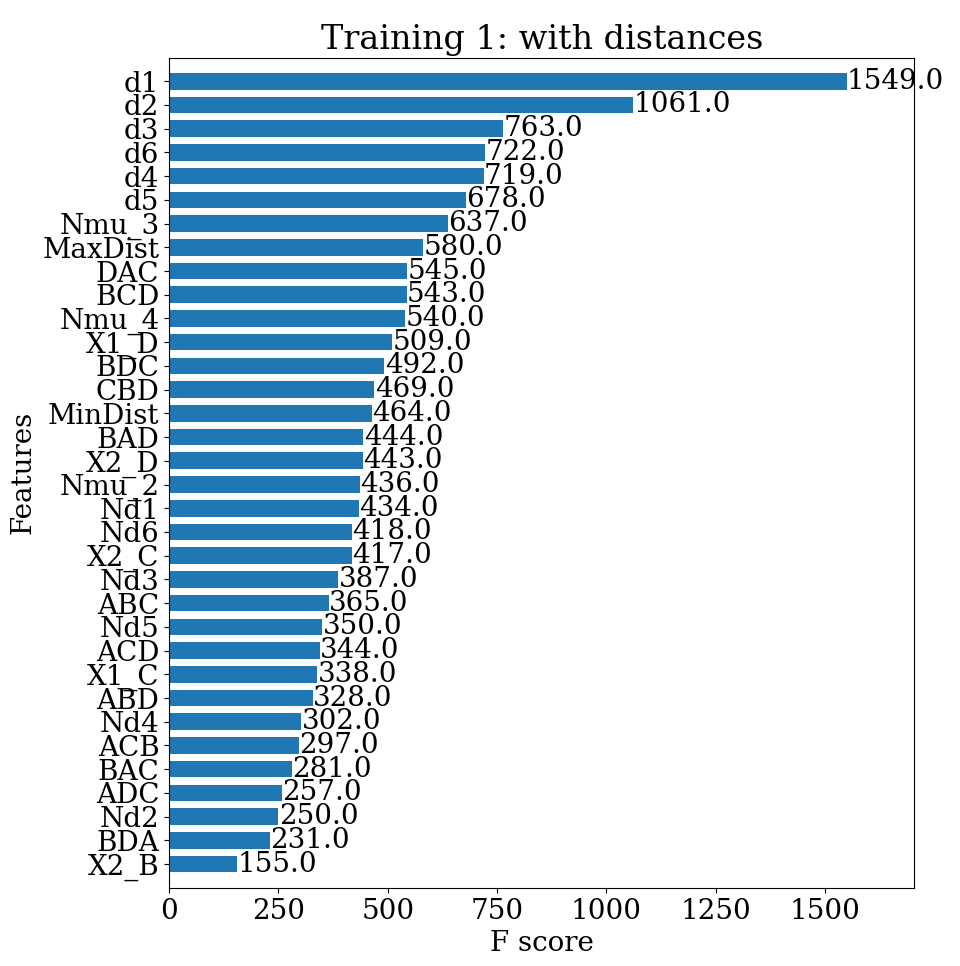}
    \includegraphics[width=0.42\textwidth]{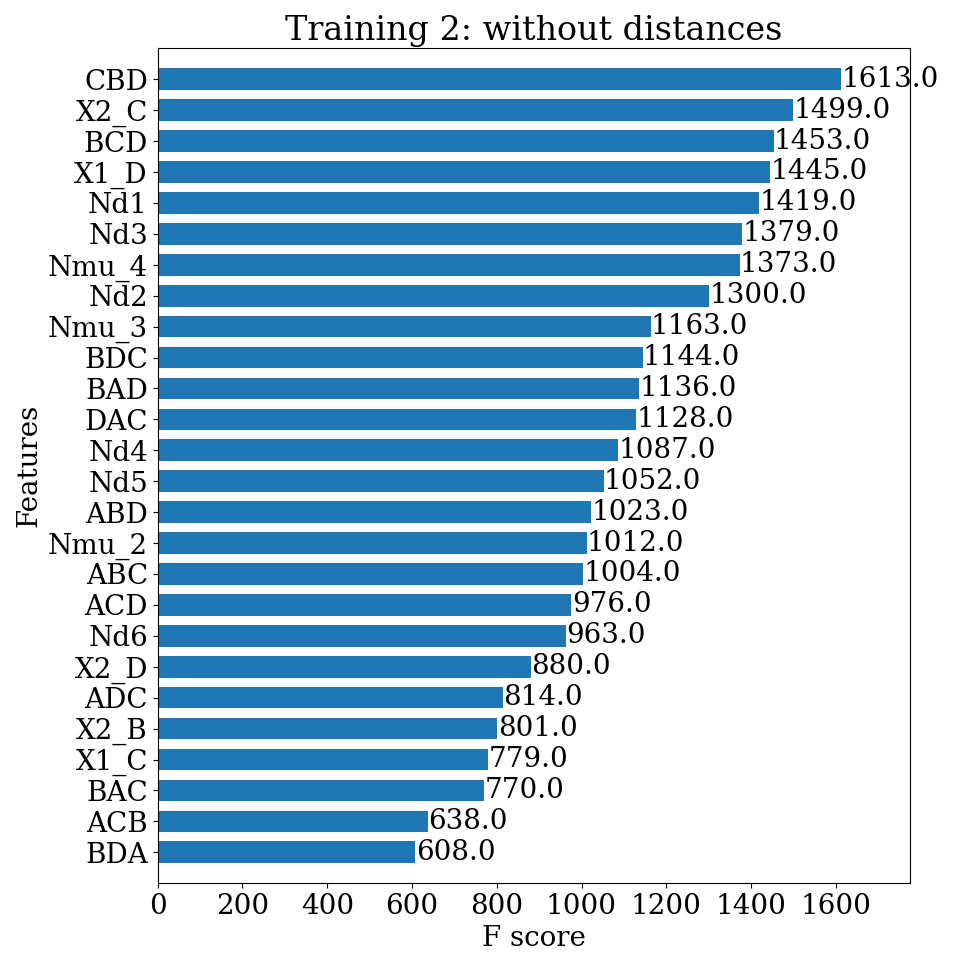}
    \caption{Feature importance of the XGBoost model trained on the first set of parameters from Table \ref{table_parametres} (left) and on all parameters (right). The training hyperparameters are 10, 0.1, and 50, respectively, for the maximum depth of trees, the learning rate, and the maximum number of trees. Features are ordered from top to bottom in order of decreasing importance.}
    \label{figure_importance}
\end{figure*}

Figure \ref{figure_importance} presents the importance of the features (F score) for the two training sessions of the XGBoost model. The feature importance analysis revealed that the most crucial parameters in training(dist) are the distances. These parameters are important in the classification because many of our gravitational lens simulations are compact configurations, due to the lack of abundant massive halos in the EAGLE simulation we analysed and because there are relatively few compact configurations of stars in our training set as a result of the selection made by \cite{2019Delchambre}. When these distances are removed, training(basic) session, the parameters that become important for classification are the angles and positions of sources C and D. 
To address the issue of the predominance of distance in the classification, we decided to conduct two separate training sessions and compare their results.

\subsection{Performances}
The optimisation phase of the XGBoost hyperparameters was carried out through an iterative process applied to 80\% of the multiplets in our training catalogue using the GridSearchCrossValidation method. The hyperparameter grid used in this search is detailed in Table~\ref{tab:xgboost_hyperparameters}, which encompasses a comprehensive range of values for key parameters such as learning rate, max depth, number of estimators, subsample, and colsample by tree. Based on the results of the GridSearchCrossValidation process, the selected hyperparameters for the XGBoost model were: \texttt{learning\_rate} = 0.1, \texttt{max\_depth} = 15, \texttt{n\_estimators} = 50, \texttt{subsample} = 1.0, and \texttt{colsample\_bytree} = 0.8, which provided optimal performance on the training set. Once these hyperparameters were determined for both training sessions, the performance of the two models was tested on the remaining 20\% of the multiplets not used in the training.

\begin{table}[h]
\centering
\caption{Hyperparameters optimised through grid search.}
\label{tab:xgboost_hyperparameters}
\begin{tabular}{lc}
\hline
Hyperparameter & Values \\
\hline
Learning Rate & 0.01, 0.1, 0.3 \\
Max Depth & 7, 10, 15, 20 \\
Number of Estimators & 5, 10, 25, 50, 100 \\
Subsample & 0.5, 0.8, 1.0 \\
Colsample by Tree & 0.5, 0.8, 1.0 \\
\hline
\end{tabular}
\tablefoot{
(1) Learning Rate: Controls step size during boosting, thus determining the contribution of each tree; smaller values require more trees but can improve model generalisation. (2) Max Depth: Maximum tree depth limiting model complexity and preventing overfitting; deeper trees capture more intricate patterns but risk memorisation. (3) Number of Estimators: Total number of trees constructed in the ensemble; more trees can improve predictive performance but also increase computational complexity. (4) Subsample: Fraction of training data used in each tree; this introduces randomness and potentially reduces overfitting; values < 1.0 create stochastic gradient boosting. (5) Colsample by Tree: Proportion of features randomly selected when constructing each tree; this promotes feature diversity and reduces the correlation between trees.
}
\end{table}

We present in Table \ref{conf} the true positive rate (TPR), the true negative rate (TPN), the false positive rate (FPR), and the false negative rate (FNR) regarding the prediction of `lens' and `group of stars' classes for the two trainings. These quantities are important metrics to qualify the performance of the trainings. 

\noindent $TPR = \frac{TP}{TP+FN}$, $TNR = \frac{TN}{TN+FP}$, $FPR = \frac{FP}{FP+TN}$, $FNR = \frac{FN}{FN+TN}$.\\
\\
Here, TP (true positives) is the number of correctly identified positive instances, TN (true negatives) is the number of correctly identified negative instances, FP is the number of negative instances incorrectly identified as positive, and FN is the number of positive instances incorrectly identified as negative.
The TPR measures the proportion of actual `lenses' that are correctly identified by the model. 
The TNR measures the proportion of actual `groups of stars' that are correctly identified by the model. 
The FPR measures the proportion of actual `groups of stars' that are incorrectly identified as `lenses' by the model.
The FNR measures the proportion of actual `lenses' that are incorrectly identified as `groups of stars' by the model.

\begin{table}
\centering
\caption{Performance parameters.}
\begin{tabular}{l|cll}
\hline
Actual class &\multicolumn{3}{c}{Predicted class rates (\%)} \\
\hline
\hline
\multicolumn{4}{c}{Training(dist)} \\
\hline
                   & Lens      & group of stars \\
\hline
 Lens              & TPR= 99.99    & FNR= 0.04 \\
\hline
 group of stars    & FPR= 0.007     & TNR= 99.96   \\
\hline
\hline
\multicolumn{4}{c}{Training(basic)} \\
\hline
                  & Lens      & group of stars \\
\hline
 Lens              & TPR= 99.99    & FNR= 0.16 \\
\hline
 group of stars    & FPR= 0.004   & TNR= 99.84    \\
\hline
\end{tabular}
\label{conf}
\tablefoot{
The TPR, TNR, FNR, and FPR in the prediction of 'Lens' and 'group of stars' classes for training(dist) and training(basic).
}
\end{table}

Both training sessions managed to classify lenses very well. Only a moderate number of lenses were placed in the 'star group' class, and a very low number of stars were classified as 'lens'. Training(basic) performed less well than training(dist), placing 0.24\% of lenses in the 'star group' class. The FPR and FNR rates (misclassified objects) for real cases are expected to be higher since micro-lensing, which affects both the geometry and the fluxes in lens simulations, is not accounted for in our simulations.
Based on these results, we leaned towards adopting training(dist) as the preferred model. However, as mentioned earlier, our simulations in the training catalogue under-represent large lenses, leading to a classification that is dependent on distance (compact configurations are more likely to be interpreted as lenses, while larger configurations are rejected more as compatible with groups of stars). Therefore, we maintained both training models when moving forward and compared their scores to select the best lens candidates.

\subsection{Validation}
\begin{figure}[!h]
\centering
    \includegraphics[width=0.49\textwidth]{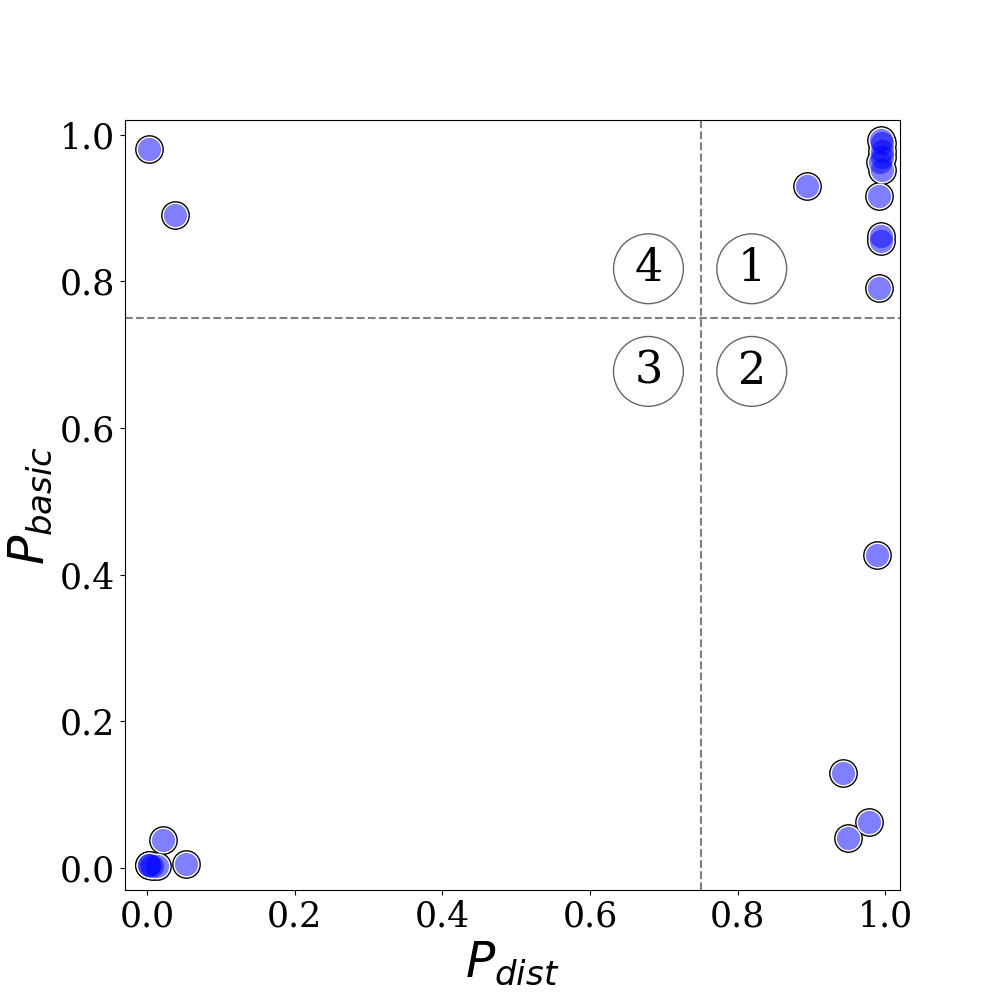}
    \caption{Probability score P$_{\textrm{basic}}$ to be a lens along with the probability score P$_{\textrm{dist}}$ for known gravitational lenses. Dotted lines separate four quadrants.}
    \label{proba_known_quads}
\end{figure}

As our aim is to assess the efficiency of our two models in classifying lenses under real conditions, we used our two models to classify 24 spectroscopically confirmed quads from \cite{ 2018Ducourant} with \textit{Gaia} measurements for their four images. Some known lenses unfortunately have only three images and are therefore not included in the analysis. Figure \ref{proba_known_quads} compares the two probability scores obtained by XGBoost for these 24 quads.

We first observed that 11 quads (46\%) -- 2MASXJ01471020+4630433 \citep{2017Berghea}, GraL024848742+191330571 \citep{2019Delchambre}, WISE025942.9-163543 \citep{2018Schechter}, HE0435-1223 \citep{2002Wisotzki}, GRAL080357714+390823333 \citep{2024Jalan}, RXJ0911+0551 \citep{1997Bade}, PG1115+080 \citep{1980Weymann}, SDSS1138+0314 \citep{2006Eigenbrod}, H1413+117 \citep{1988Magain}, GraL1537-3010 \citep{2019Delchambre,2019Lemon}, WFI2033-4723 \citep{2004Morgan} -- are accurately identified by both models with scores greater than 0.75 (quadrant 1). The quads tend to be compact configurations with separations smaller than 3".
Five quads -- WG0214-2105 \citep{2019Spiniello}, RXJ1131-1231 \citep{2003Sluse}, B1422+231 \citep{1992Patnaik}, J1606-2333 \citep{2018Lemon}, and J2145+6345 \citep{2019Lemon} -- are identified by model(dist) but rejected by model(basic), quadrant 2.

In quadrant 3, both models reject six quads (including the emblematic Einstein cross G2237+0305) with very low scores (<0.1): GraL065904.1+162909 \citep{2021Stern}, 2MASSJ11344050-2103230 \citep{2018Lucey}, 2MASSJ13102005-1714579 \citep{2018Lucey}, J1606-2333 \citep{2018Lemon}, GraL203802-400815 \citep{2018Krone-Martins}, and G2237+0305 \citep{1985Huchra}. Several factors contribute to these configurations being poorly recognised. Three of them lack compactness, with the maximum angular separations generally being higher than 4". They exhibit significant elongation, thus deviating from the configurations produced by our SIE plus shear gravitational lens simulation model. For the Einstein cross Q2237, the impact of dust in image D was estimated by \cite{2008Eigenbrod}, highlighting a critical consideration in the analysis of gravitational lenses. While dust is not typically a critical factor for most gravitational lens systems, Q2237 presents a notable exception due to its specific galactic structure. GravLens likely encountered challenges in accurately measuring lens components and their luminosities, partly due to the preponderance of the surrounding deflecting galaxy and the complex dust distribution, which significantly impacts microlensing phenomena.

The two quads -- GraL081828.3-26132 \citep{2021Stern} and J1721+8842 \citep{2018Lemon} -- in quadrant 4 that the model(basic) identified securely (P$_{\textrm{basic}}$>0.75) and that model(dist) classes as a group of stars (P$_{\textrm{dist}}$<0.10) are large configurations with a MaxDist greater than 4". This is a typical consequence of the under-representation of large quad configurations in our training set.

We observed that both models perform quite well on typical compact configurations but diverge when both the complexity and the size of the configurations increase. Model(dist) successfully classifies 67\% of the quads (quadrant 1 plus quadrant 2), and model(basic) is successful in 54\% of the cases (quadrant 1 plus quadrant 4). It is clear that model(dist) performs better than model(basic), but model(basic) slightly outperforms when identifying large configurations.  

If we consider the two scores above 0.75 together, it is possible to identify 18 of the 24 quads analysed (75\%; quadrants 1, 2, and 4), which is a very good performance when considering that the model used to produce lens simulations for the training set does not account for micro-lensing effects or multiple deflectors. The limitation of our current methodology is primarily linked to the simplicity of the SIE model plus shear for a certain proportion of known quads, and it is also due to the under-representation of large lenses in our training set. 

\section{Application to the GravLens catalogue}\label{sec:applications}

\subsection{Selection of quadruplets}
We applied our algorithm to the 81\,576 multiplets of the GravLens catalogue, each of which contain four or more sources. When there were more than four sources, all combinations of the sources within the multiplet were considered. We ended up with 1\,128\,000 quadruplets to analyse. The sources in the quadruplets were then ranked with respect to their magnitude so that the brightest was identified as A and the faintest as D.

 Before applying XGBoost to the multiplets, we filtered out configurations that are obviously non-lens. Indeed, in a four-image lensed quasar, it is impossible for one of the images to be contained within the triangle formed by the other three images. Among the \textit{Gaia} quadruplets, we rejected 225\,761 such cases. This constraint enabled us to eliminate 20\% of the multiplets that do not meet this criterion and left us with 902\,239 quadruplets to analyse, corresponding to 65\,996 multiplets from GravLens.

\subsection{Classification}
\begin{figure}[!h]
\centering
    \includegraphics[width=0.49\textwidth]{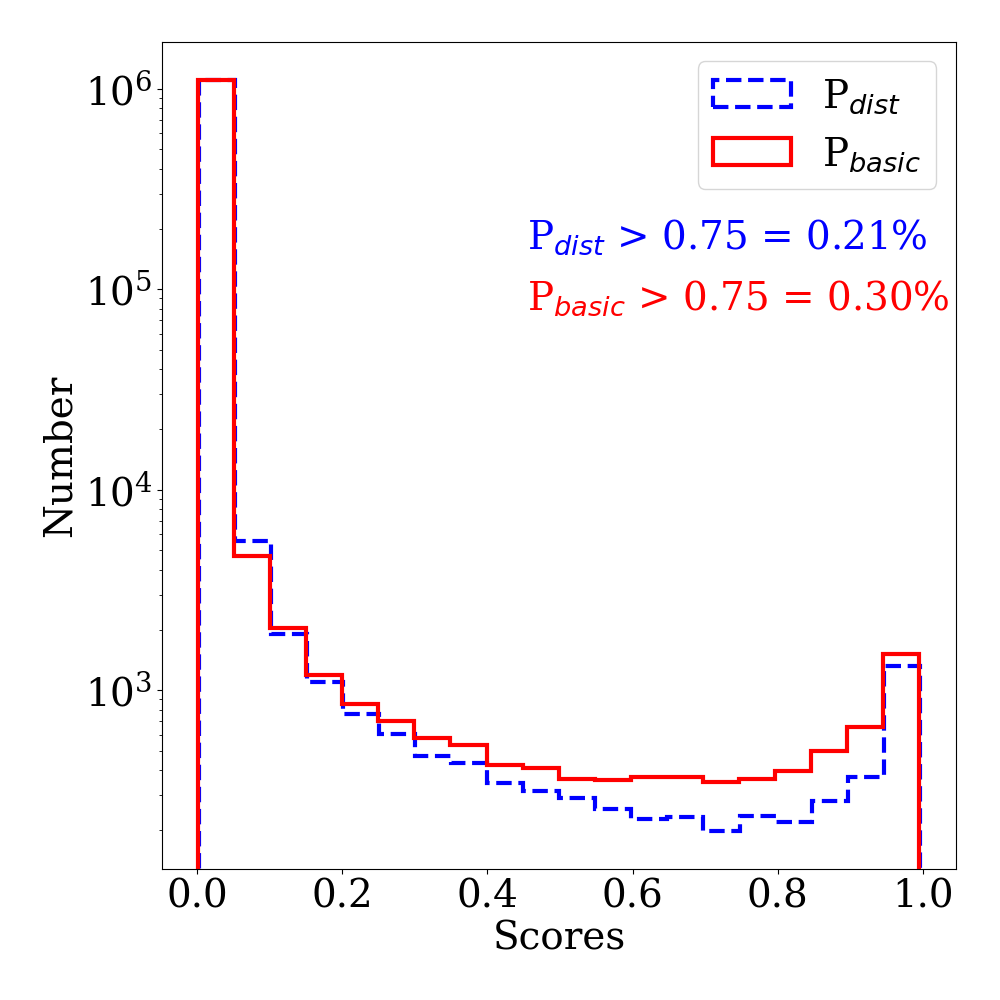}
    \caption{Scores for the 902\,239 GravLens quadruplets analysed.}
    \label{scores_fpr}
\end{figure}

We applied our classification algorithm (both training models) to the remaining 902\,239 quadruplets and obtained two scores for each: P$_{\textrm{dist}}$ (training with distances) and P$_{\textrm{basic}}$ (training without distances).
Figure \ref{scores_fpr} presents the distribution of both scores for all quadruplets analysed. We observed that most multiplets have low scores with both models, consistent with GravLens containing a majority of non-lens objects. Model(dist) is more selective than model(basic), which assigns high scores to a smaller population of multiplets.

\begin{figure}[!h]
\centering
    \includegraphics[width=0.49\textwidth]{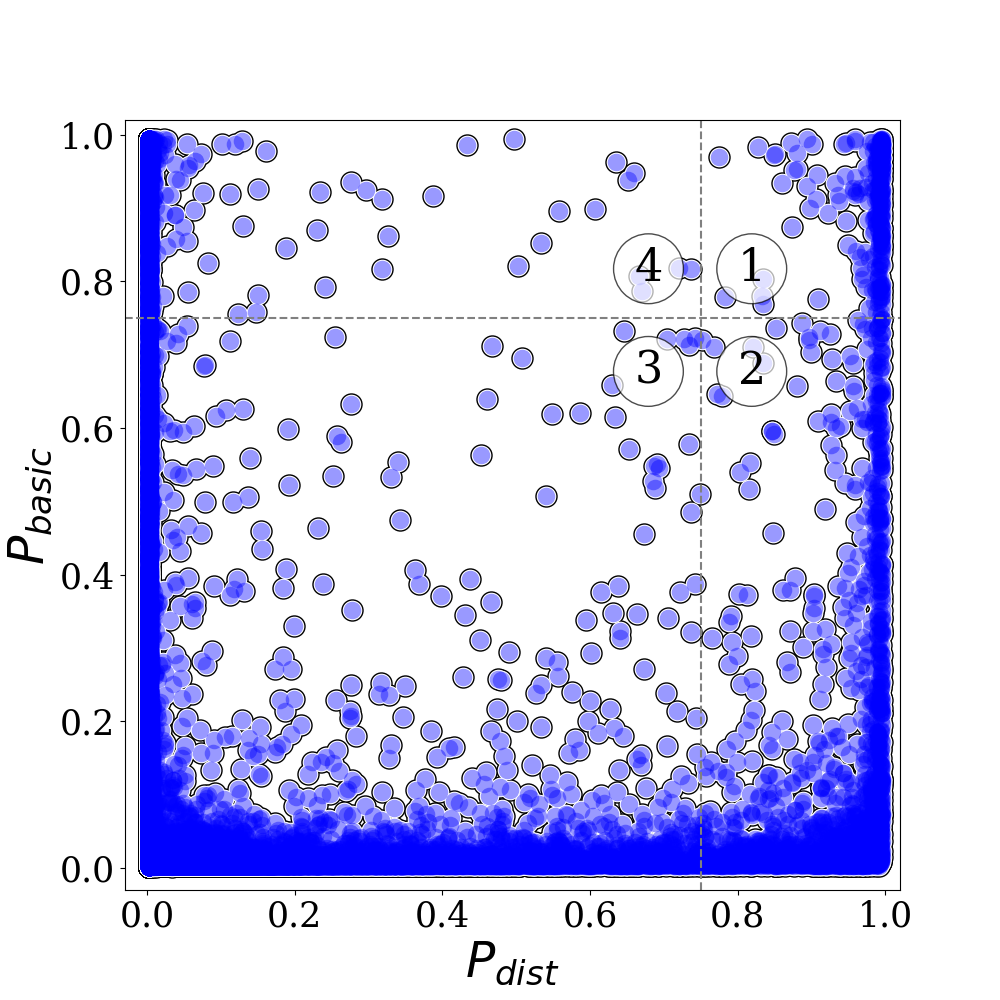}
    \caption{Probability score P$_{\textrm{basic}}$ for being a lens plotted against probability score P$_{\textrm{dist}}$ for 902,239 quadruplets from the GravLens catalogue. The plot is divided into four quadrants with a threshold of 0.75 for both axes, creating regions defined by their probability score combinations.}
    \label{scores_fpr_P1_P2}
\end{figure}

We present a comparison of P$_{\textrm{dist}}$ and P$_{\textrm{basic}}$ in Fig. \ref{scores_fpr_P1_P2}, and the counts of quadruplets in the various quadrants are shown in Table \ref{fpr_quadrants}. In this table, we also indicate the number of multiplets involved (multiplets with more than four sources correspond to more than one quadruplet).

\begin{table}
    \centering
    \caption{Number of GravLens quadruplets (configurations with all possible combinations of four sources in the field of view) and multiplets (locations with at least four sources) with P${\textrm{dist}}$ and P${\textrm{basic}}$ placing them in the various quadrants of Fig. \ref{scores_fpr_P1_P2}.}
     \begin{tabular}{crr}
     \hline 
         Quadrant    &  No of quadruplets   & No of multiplets\\
     \hline
            1       &   226             &   201     \\
            2       &   1847            &   798     \\
            3       &   897\,369        &   62\,662 \\
            4       &   3\,141          &   2\,221  \\  
     \hline
    \end{tabular}
    \label{fpr_quadrants}
\end{table}
As expected, most quadruplets fall into quadrant 3, where both models reject them because they are identified as being stars. We note that we expected at most a few hundred quads in the \textit{Gaia} catalogue \citep{2016Finet}. A moderate number of quadruplets lie in quadrants 2 and 4 (these are interesting sources to investigate further). Finally, 226 quadruplets lie in quadrant 1, where both models identify them as lenses. These are the best candidates.

\begin{figure}[!h]
\centering
    \includegraphics[width=0.49\textwidth]{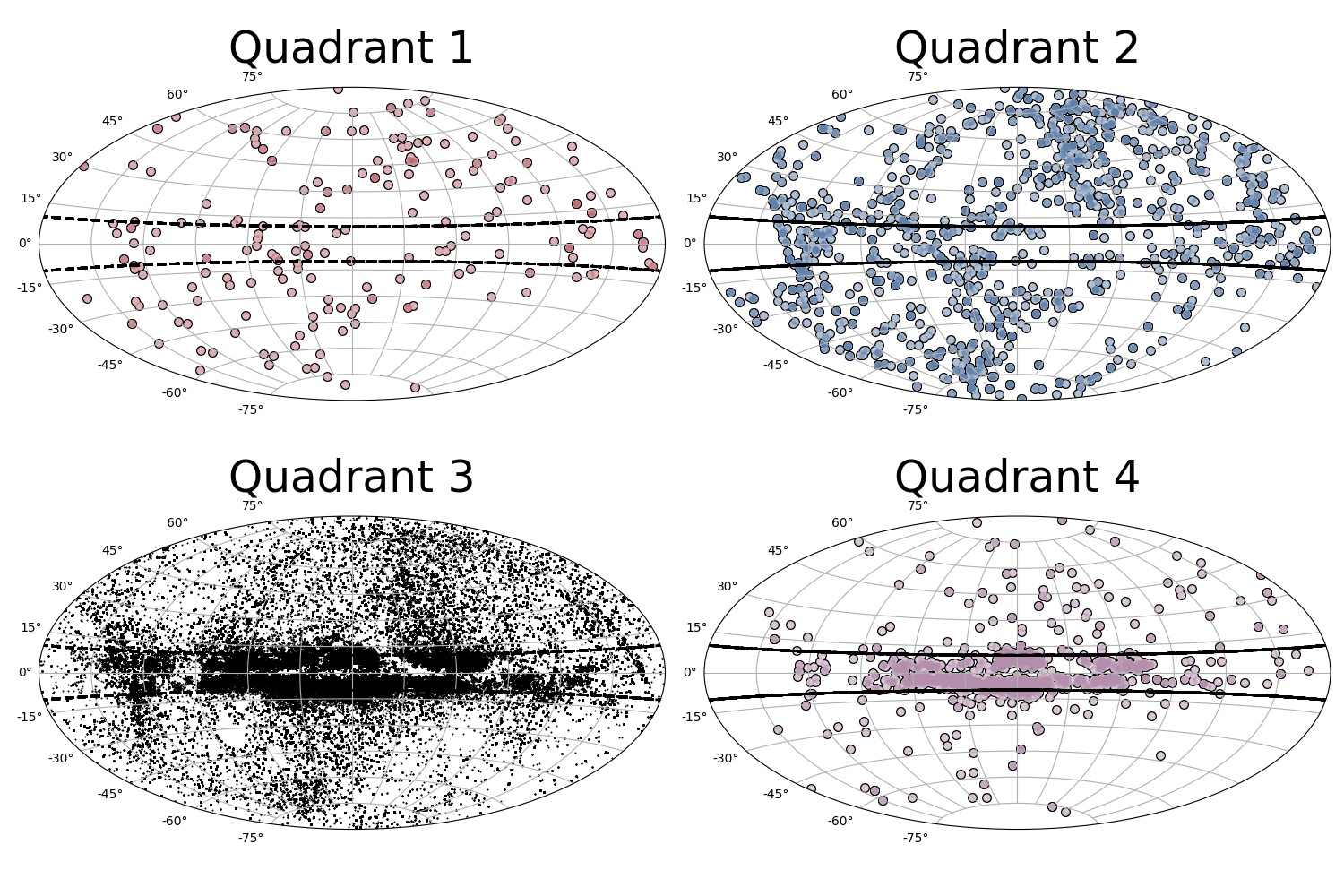}
    \caption{Sky distribution in galactic coordinates of the quadruplets of each quadrant of Fig. \ref{scores_fpr_P1_P2}. Dotted lines correspond to |b|=10.}
    \label{quadrant_lb}
\end{figure}

To further analyse the quadruplets in each quadrant, we examined their sky distribution in galactic coordinates (Fig. \ref{quadrant_lb}). We note that the spatial distribution of sources from quadrants 3 and 4 is heterogeneous and has a very high density in the galactic plane (|galactic\_lat|<10°), suggesting that these sources are most likely quadruplets of stars that correctly replicate a lens configuration. In contrast, sources from quadrants 1 and 2 exhibit a more homogeneous sky coverage. 

To compile a list of candidates for the spectroscopic follow-ups we are planning, we selected the 201 multiplets from quadrant 1 with good probability scores ($P_{basic}$ and $P_{dist}$ > 0.8). Among them, we further refined our selection based on galactic latitude (|b| > 15°) and performed a visual inspection, identifying the 48 most promising candidates. The final list of these top candidates is presented in Appendix~\ref{appendix1}.

However, we are completely aware that some sources of GravLens are issued from the fragmentation of single galaxies by GravLens into multiple sources that can mimic lens configurations. Further filtering and visual inspection are mandatory to reject this type of contaminant. 

\textit{Gaia}'s limiting magnitude (approximately G=21 mag) severely limits the number of quads for which the space observatory can detect all four images. For pragmatic reasons, we limited our study to multiplets of four images. This is why future development of the work presented here must include analysis of triplets of sources since one source of the quads may not be detected.

\section{Conclusion}\label{sec:conclusion}

 We have presented a new method based on the XGBoost algorithm for the supervised classification of quadruply imaged quasars in large catalogues. We applied our method to the \textit{Gaia} FPR GravLens catalogue \citep{2023GaiaFPR}, which provides the celestial coordinates and G magnitudes of all sources detected by the satellite within a radius of 6\arcsec around approximately three million known quasars.
 
To train the XGBoost algorithm, we considered two classes of sources: lensed quasars and groups of stars. For the first class, we developed a set of realistic simulations of gravitational lenses by placing a sample of quasars drawn from the Milliquas catalogue behind galaxy halos measured in the EAGLE set of cosmological simulations. We used SIE plus shear to model these deflectors. The second class of the training set comprises quadruplets of stars selected from the \textit{Gaia} catalogue.
We carried out two XGBoost training sessions: one considering all parameters available, including distances, and a second without the distances between sources. This training approach resulted in two separate scores.

We succeeded in building a parameter space that appears to be efficient for quasar lens classification. The discriminant parameters describe aligned and normalised configurations. This parameter space exhibits distinct regions corresponding to specific gravitational lens configurations.

Analysis of the feature importance shows that besides the distances that are preponderant in training using distance parameters, the three angles of the BCD triangle defined by the three faintest sources of the multiples are important and so is the location of image D in this parameter space. The flux ratios of images C and D relative to image A are also important for the classification. The performances, as measured by the rate of correctly classified lenses and group of stars, are above 99.99 and 99.84, respectively. These results are satisfactory in terms of completeness but less so in terms of purity.

We applied our trained algorithm to the 902\,239 selected quadruplets of sources in the \textit{Gaia} FPR GravLens catalogue and calculated the two scores for each multiplet. By comparing the scores obtained, we selected a pool of 1\,127 multiplets with at least one score larger than 0.75. From these, 201 have both scores above 0.75 and are excellent candidates. We are currently examining these candidates one by one to further assess their nature.

The work presented here is focused on setting up the method and selecting the discriminating training parameters to produce a tool that robustly classifies multiplets of sources. To go further, one can improve the training set and the parameters used for the classification. For the training set, we can improve the lens simulations by incorporating micro-lensing effects. Other parameters could also be used during the training, such as colours, galactic coordinates, star density in the region, and astrometric parameters (e.g. parallax, proper motion, etc.). These parameters are not available in the \textit{Gaia} FPR catalogue but can be extracted from other catalogues for a large set of our sources. 

The GravLens catalogue also contains more than 234\,000 multiplets with three sources. These sources certainly include quads where one of the images was not detected by the satellite (e.g. eight known quads are present among these triplets). In order to be able to analyse them, simulations of triply imaged quasars should be added to the training set by removing the faintest image of the multiples and the classifier adapted to this case. 

We find the current results all the more encouraging given that microlensing is not accounted for per se. Nevertheless, the ground-based spectroscopic monitoring campaigns we are continuing will enable us to determine the real performance of this new tool.

\begin{acknowledgements}

We acknowledge the french national program PN-GRAM and Action Sp\'ecifique \textit{Gaia} as well as Observatoire Aquitain des Sciences de l'Univers (OASU) for financial support along the years. \\
Our work was eased by the use of the data handling and visualisation software TOPCAT \citep{2005Taylor}.
This research has made use of "Aladin sky atlas" developed at CDS, Strasbourg Observatory, France \citep{Aladin2014ASPC..485..277B, Aladin2000A&AS..143...33B}.
This research has made use of the VizieR catalogue access tool, CDS, Strasbourg, France.
This work has made use of data from the European Space Agency (ESA) mission
{\it Gaia} (\url{https://www.cosmos.esa.int/gaia}), processed by the {\it Gaia}
Data Processing and Analysis Consortium (DPAC, \url{https://www.cosmos.esa.int/web/gaia/dpac/consortium}). Funding for the DPAC
has been provided by national institutions, in particular the institutions
participating in the {\it Gaia} Multilateral Agreement.

\end{acknowledgements}

\bibliographystyle{aa}
\bibliography{bibliography}

\begin{appendix}
\onecolumn

\section{Lens candidates}\label{appendix1}

\begin{figure*}[!h]
\centering
    \includegraphics[width=0.32\textwidth]{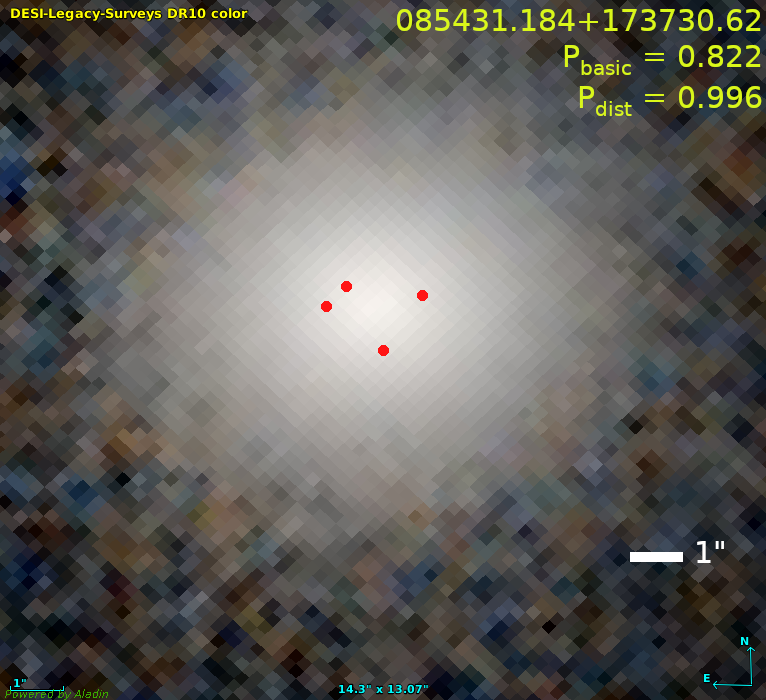}
    \includegraphics[width=0.32\textwidth]{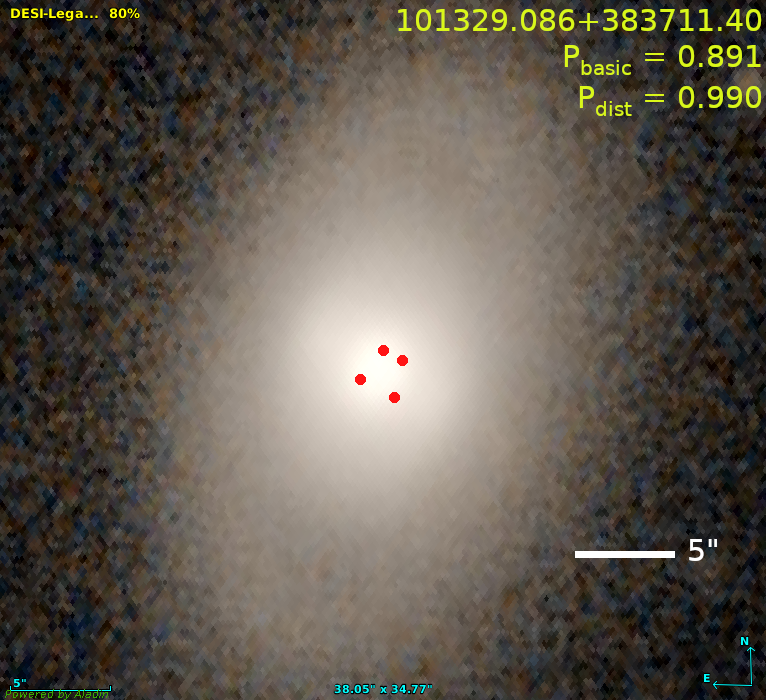}
    \includegraphics[width=0.32\textwidth]{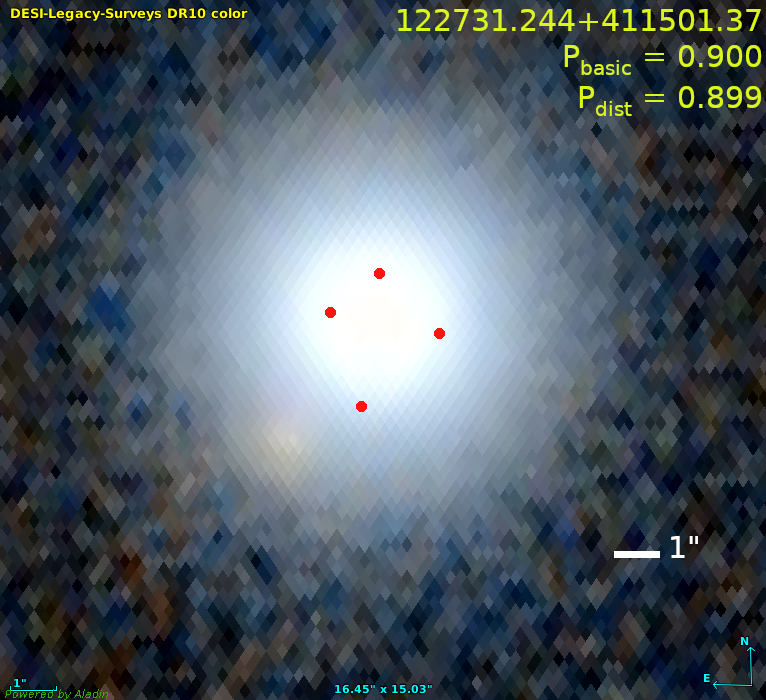}
    \includegraphics[width=0.32\textwidth]{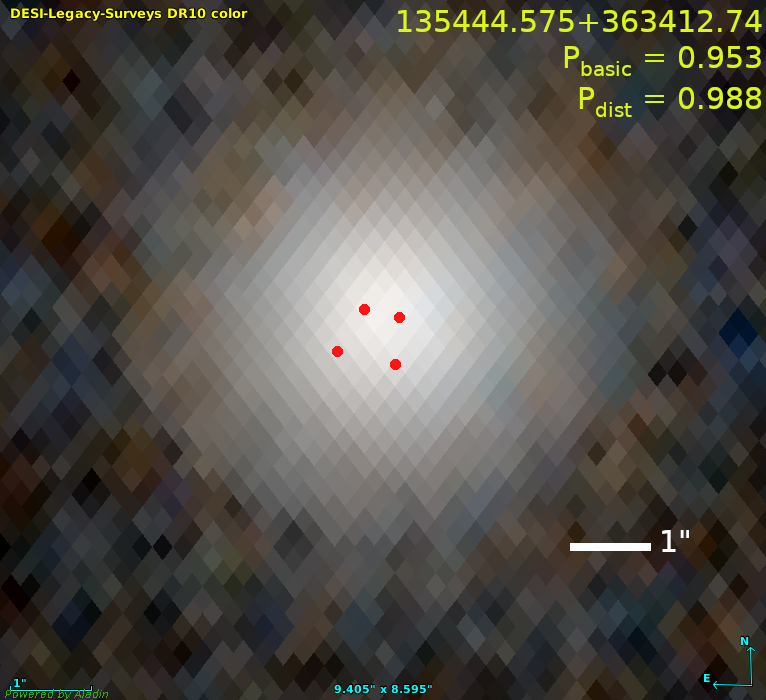}
    \includegraphics[width=0.32\textwidth]{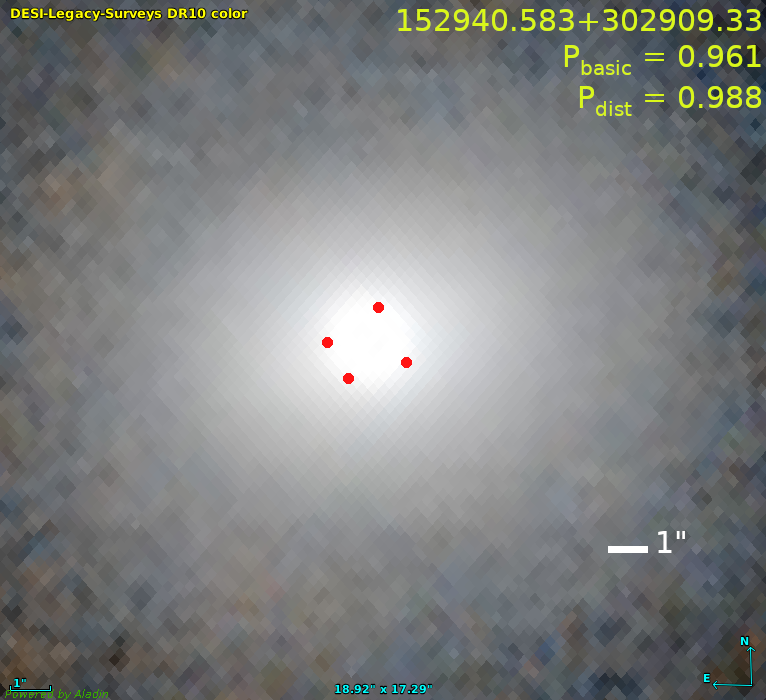}
    \includegraphics[width=0.32\textwidth]{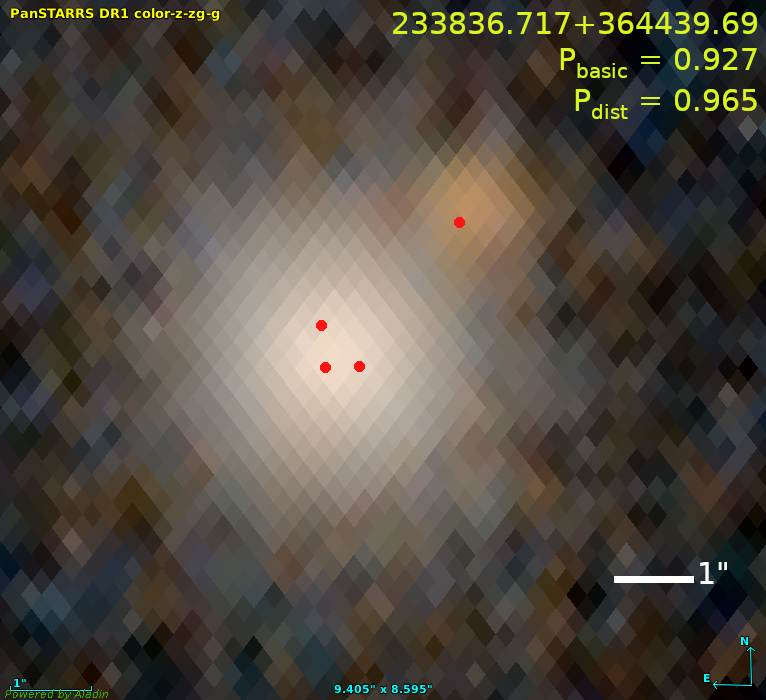}
    \caption{Mosaic of cutouts from the DESI Legacy Imaging Surveys \citep{2019Dey} and Pan-STARRS1 \citep{2018Chambers} imaging showing promising gravitational lens candidates. Each panel displays the system identifier and includes the two probability scores ($P_{basic}$, $P_{dist}$). The FPR GravLens sources are superimposed on each image.}
    \label{mosaic}
\end{figure*}

This section presents mosaic cutouts and a list of our most promising lens candidates using imaging data from the DESI Legacy Imaging Surveys \citep{2019Dey} and Pan-STARRS1 \citep{2018Chambers}. The selected cutouts (Fig. \ref{mosaic}) showcase the spatial configuration of the components, allowing for visual inspection of their morphology and relative positions.

The list of candidates (Table \ref{list-candidate}) only shows the most promising candidates. These candidates have been selected by keeping only those candidates whose galactic latitude is greater than 15 degrees and whose $P_{basic}$ and $P_{dist}$ scores are greater than 0.8. In addition, a visual inspection allowed us to retain only the most promising candidates.

\begin{table*}
\centering
    \caption{Most promising gravitational lens candidates.}
     \begin{tabular}{lccl}
     \hline
           GravLensName                & $P_{basic}$ & $P_{dist}$ \\
     \hline
           DR3Gaia002250.874-341602.72 &  0.906 & 0.996 \\
           DR3Gaia003612.327-464300.50 &  0.971 & 0.996 \\
           DR3Gaia005936.699-134725.04 &  0.828 & 0.993 \\
           DR3Gaia021838.362-450442.80 &  0.957 & 0.996 \\
           DR3Gaia022632.876-514803.92 &  0.925 & 0.994 \\
           DR3Gaia033647.500-344422.73 &  0.976 & 0.991 \\
           DR3Gaia034555.352-571147.71 &  0.990 & 0.996 \\
           DR3Gaia035400.850-720801.70 &  0.839 & 0.972 \\
           DR3Gaia040106.637-160639.01 &  0.885 & 0.992 \\
           DR3Gaia045106.893-395138.51 &  0.988 & 0.996 \\
           DR3Gaia045711.543-360819.61 &  0.968 & 0.995 \\
           DR3Gaia050209.047+033149.93 &  0.842 & 0.995 \\
           DR3Gaia053204.686-201534.13 &  0.844 & 0.990 \\
           DR3Gaia060236.534-535600.05 &  0.841 & 0.996 \\
           DR3Gaia060712.719+683319.44 &  0.901 & 0.995 \\
           DR3Gaia064713.748-530756.10 &  0.888 & 0.994 \\
           DR3Gaia074738.332-732553.15 &  0.959 & 0.995 \\
           DR3Gaia082940.765-713749.77 &  0.870 & 0.991 \\
           DR3Gaia085431.184+173730.62 &  0.822 & 0.996 \\
           DR3Gaia094632.029+351949.40 &  0.887 & 0.996 \\
           DR3Gaia101329.086+383711.40 &  0.891 & 0.990 \\
           DR3Gaia104114.659-830859.87 &  0.917 & 0.994 \\
           DR3Gaia110057.151+103028.69 &  0.973 & 0.994 \\
           DR3Gaia110504.207+505949.94 &  0.893 & 0.993 \\
           DR3Gaia111441.792+574931.32 &  0.817 & 0.991 \\
           DR3Gaia111832.300+534852.24 &  0.903 & 0.991 \\
           DR3Gaia114338.825-013845.08 &  0.931 & 0.993 \\
           DR3Gaia122731.244+411501.37 &  0.900 & 0.899 \\
           DR3Gaia124127.947-162922.05 &  0.985 & 0.995 \\
           DR3Gaia130550.130+401700.44 &  0.951 & 0.996 \\
           DR3Gaia131131.004+462030.32 &  0.969 & 0.996 \\
           DR3Gaia133815.871+043233.53 &  0.984 & 0.990 \\
           DR3Gaia134244.428+350346.45 &  0.927 & 0.995 \\
           DR3Gaia135444.575+363412.74 &  0.953 & 0.988 \\
           DR3Gaia135534.320+594434.03 &  0.986 & 0.995 \\
           DR3Gaia143208.703-270432.18 &  0.805 & 0.990 \\
           DR3Gaia150720.640+301529.74 &  0.835 & 0.995 \\
           DR3Gaia151638.719+410148.75 &  0.940 & 0.988 \\
           DR3Gaia152940.583+302909.33 &  0.961 & 0.988 \\
           DR3Gaia153016.153+270551.01 &  0.978 & 0.996 \\
           DR3Gaia155719.490+281354.48 &  0.874 & 0.873 \\
           DR3Gaia160653.603+034736.58 &  0.992 & 0.995 \\
           DR3Gaia180416.050+644414.80 &  0.880 & 0.993 \\
           DR3Gaia205440.111-420426.30 &  0.970 & 0.993 \\
           DR3Gaia210909.970-094014.75 &  0.848 & 0.995 \\
           DR3Gaia212554.239-400037.32 &  0.846 & 0.995 \\
           DR3Gaia233029.711-631831.17 &  0.850 & 0.996 \\
           DR3Gaia233836.717+364439.69 &  0.927 & 0.965 \\
     \hline
    \end{tabular}
    \tablefoot{
     All candidates have a galactic latitude |b| > 15° and probability scores $P_{basic}$ and $P_{dist}$ > 0.8. For each candidate, we provide: the \textit{Gaia} FPR GravLensName (DR3GaiaHHMMSS.sss+DDMMSS.ss) and the probability scores from our selection pipeline.
    }
    \label{list-candidate}
\end{table*}

\end{appendix}

\end{document}